%% file: PAPER-TopDiffXS8TeV.tex
\pdfoutput=1
\newcommand*{\ATLASLATEXPATH}{}
\documentclass[UKenglish,texlive=2011,cernpreprint, subfigure=true,txfonts=true]{\ATLASLATEXPATH atlasdoc}

\usepackage[utf8]{inputenc} 
\usepackage{a4wide}
\usepackage{\ATLASLATEXPATH atlaspackage}
\usepackage{\ATLASLATEXPATH atlasbiblatex}

\usepackage{\ATLASLATEXPATH atlascontribute}

\usepackage{\ATLASLATEXPATH atlasphysics}

\graphicspath{{logos/}{figures/}}
\usepackage[utf8]{inputenc}

\usepackage{PAPER-TopDiffXS8TeV-defs}

\input{PAPER-TopDiffXS8TeV-metadata}
\hypersetup{pdftitle={ATLAS draft},pdfauthor={The ATLAS Collaboration}}

\begin{document}

\maketitle

\tableofcontents
\input{introduction}
\input{detector}
\input{datamcsamples}

\input{objdefinitions}

\input{backgrounds}

\input{systematics}

\input{datamccomp}

\input{unfolding}

\input{results}
\input{conclusions}

\section*{Acknowledgments}
\input{Acknowledgements}
\clearpage
\appendix
\input{systBreakDown}

\printbibliography

\clearpage
\input{atlas_authlist}

\end{document}

%% file: PAPER-TopDiffXS8TeV-metadata.tex
\AtlasTitle{Measurement of the differential cross-section of highly boosted top quarks as a function of their
transverse momentum in $\sqrt{s}$ = 8 \TeV\
proton--proton collisions using the ATLAS detector}

\author{The ATLAS Collaboration}

\AtlasRefCode{TOPQ-2014-15}

\PreprintIdNumber{CERN-PH-EP-2015-237}

\AtlasJournal{Phys.\ Rev.\ D.}

\AtlasAbstract{%
The differential cross-section for pair production of top quarks with
high transverse momentum is measured in 20.3~fb$^{-1}$ of proton--proton collisions at a
center-of-mass energy of 8 \TeV. The measurement is performed for
$t\bar{t}$ events in the lepton+jets channel. The cross-section is
reported as a function of the hadronically decaying top quark
transverse momentum for values above 300 \GeV. The hadronically decaying
top quark is reconstructed as an anti-$k_t$ jet with radius parameter
$R=1.0$ and identified with jet substructure techniques.
The observed yield is corrected for detector effects
to obtain a cross-section at particle level in a fiducial region close to the event 
selection. A parton-level cross-section
extrapolated to the full phase space is also reported for top quarks with 
transverse momentum above 300 \GeV.
The predictions of a majority of next-to-leading-order
and leading-order matrix-element Monte Carlo
generators are found to agree with the measured cross-sections.
}

%% file: introduction.tex
\section{Introduction}
\label{sec:intro}
The large number of top--antitop quark ($\ttbar$) pairs produced at the LHC provide a unique opportunity to improve our understanding of $\ttbar$ production and test the Standard Model (SM) at the \TeV\ scale. 
New phenomena beyond the Standard Model may distort the top quark transverse momentum ($\pt$) spectrum, in particular at high $\pt$ (see, e.g., Refs.~\cite{Atwood:1994vm,Englert:2012by}), and could thus be revealed by a precise measurement. Moreover, due to their high cross-section at the LHC and rich experimental signature, $\ttbar$ events constitute a dominant background to a wide range of searches for new massive particles. A better understanding of the production of high-momentum top quarks, including a more precise determination of the parton distribution functions (PDF) of the proton, would be of great benefit to the broader LHC program.

 The initial measurements of $\ttbar$ production at the LHC have focused on a determination of the inclusive production cross-section. Now that the experimental uncertainties on these measurements (see, e.g., Refs.~\cite{Aad:2014kva,Chatrchyan:2012bra,Chatrchyan:2013faa}) are comparable to or lower than the uncertainties on the next-to-next-to-leading-order plus next-to-next-to-leading-logarithmic order (NNLO+NLLL) theory prediction~\cite{Cacciari:2011hy,Beneke:2011mq,Baernreuther:2012ws,Czakon:2012zr,Czakon:2012pz,Czakon:2013goa}, the interest in differential top quark cross-section measurements has gained traction. 
 Measurements of the differential cross-section as a function of the kinematics of the top quark, or the top--antitop quark pair, have been performed by
 the ATLAS~\cite{Aad:2012hg,Aad:2014zka,Aad:2015eia}
and CMS collaborations \cite{Chatrchyan:2012saa,Khachatryan:2015oqa}, where the highest measured top quark 
$\pt$ range is 350-800~\GeV~\cite{Aad:2014zka}.

In this paper a measurement using
techniques specifically designed to deal with the collimated decay topology of highly boosted top quarks is presented. In particular, the hadronic top quark decay is reconstructed as a single large-radius (large-$R$) jet. 
The selection and reconstruction are based on an algorithm developed \cite{ATL-PHYS-PUB-2010-008} and used in $\ttbar$ resonance searches \cite{Aad:2013nca,Aad:2012raa,Aad:2015fna,Chatrchyan:2012ku}
that increases the $\ttbar$ selection efficiency at high top quark $\pt$ and extends the kinematic reach into the \TeV\ range.
 This analysis utilizes the lepton+jets channel where one $W$ boson decays hadronically and the other leptonically 
to an electron or a muon, assuming each top quark decays to a $W$ boson and a $b$-quark. The cross-section is measured as a function of the hadronically decaying top quark $\pt$. 
A particle-level cross-section is measured in a kinematic region close to the detector-level selection, referred to in the following as fiducial region.
A parton-level differential cross-section is also reported as a function of the hadronically decaying top quark $\pt$, by further extrapolating to the full kinematic phase space except for a lower limit on top quark $\pt$ of 300~\gev. The measured cross-sections are compared to the predictions of several MC generators and PDF sets.

The object definition, event selection, and background determination used in this analysis follow closely the ones used in the search for $t\bar{t}$ resonances \cite{Aad:2015fna}. More details of these aspects of the measurement can be found in the corresponding reference.

%% file: detector.tex
\section{The ATLAS detector}
\label{sec:detector}

ATLAS is a multipurpose detector \cite{Aad:2008zzm} that provides nearly
full solid angle\footnote {ATLAS uses a
  right-handed coordinate system with its origin at the nominal
  interaction point (IP) in the center of the detector and the $z$-axis
  along the beam pipe. The $x$-axis points from the IP to the center of
  the LHC ring, and the $y$-axis points upward. Cylindrical coordinates
  ($r$,$\phi$) are used in the transverse plane, $\phi$ being the azimuthal angle
  around the beam pipe. The pseudorapidity is defined in terms of the
  polar angle $\theta$ as $\eta = - \ln \tan(\theta/2)$.} coverage around the interaction point. 
Charged-particle trajectories
are reconstructed
by the inner detector, which covers pseudorapidity $|\eta| <2.5$ 
and is composed of a silicon pixel detector, a silicon 
microstrip detector, and a transition radiation tracker (TRT).
The inner detector is surrounded by a solenoid that provides a 2 T magnetic
field.
Sampling calorimeters with several different
designs span the pseudorapidity range up to $|\eta| = 4.9$. 
High-granularity liquid-argon (LAr) electromagnetic (EM) calorimeters are used up
to $|\eta| = 3.2$. Hadronic calorimetry based on scintillator-tile
active material covers $|\eta| < 1.7$ while LAr technology is utilized
for hadronic calorimetry from $|\eta| = 1.5$ to $|\eta| = 4.9$. The
calorimeters are surrounded by a muon spectrometer. A magnetic field in the spectrometer 
is provided by air-core toroid magnets. Three layers of precision
gas chambers track muons up to $|\eta|$ = 2.7 and muon trigger chambers
cover  $|\eta| <2.4$.

%% file: datamcsamples.tex
\section{Data and Monte Carlo samples}
\label{sec:datamcsamples}
The cross-section is measured using data from the 2012 LHC $pp$ run at $\rts$ = 8 \TeV, which corresponds to an
integrated luminosity of $20.3 \pm 0.6$~\ifb. The luminosity was measured using
techniques similar to those described in Ref.~\cite{Aad:2013ucp} with a 
calibration of the luminosity scale derived from beam-overlap scans
performed in November 2012.
The average number of $pp$ interactions per bunch crossing (pileup) in 2012 was around 21. 
The sample was collected using the logical OR of two single-electron triggers with transverse 
momentum thresholds of 60 \GeV, lowered to 24 \GeV\ in the case of isolated electrons, and 
two single-muon triggers with transverse 
momentum thresholds of 36 \GeV, lowered to 24 \GeV\ in the case of isolated muons. 

Samples of Monte Carlo (MC) simulated events are used to characterize the detector
response and efficiency to reconstruct $\ttbar$ events, estimate systematic uncertainties, predict
the background contributions from various physics processes, 
and to compare the theoretical predictions with the measurement.
The simulated events are weighted such that the distribution of the
average number of $pp$ interactions per bunch crossing agrees with
data.  The samples were processed through the GEANT4\cite{SAMPLES-G4}
simulation of the ATLAS detector \cite{Aad:2010ah}.
For the evaluation of some systematic uncertainties, generated samples are passed to a fast 
simulation using a parameterization of the performance of the
ATLAS electromagnetic and hadronic calorimeters \cite{ATL-SOFT-PUB-2014-001}.
Simulated events are reconstructed using the same algorithms that are applied to the data. 

The nominal signal $\ttbar$ sample is generated using the {\sc Powheg} 
({\sc Powheg}-hvq patch4)
\cite{Frixione:2007vw} method, as implemented in the {\sc Powheg-Box}
generator \cite{Alioli:2010xd}, which is based on
next-to-leading-order (NLO) QCD matrix elements. The $h_\mathrm{damp}$ parameter, which effectively regulates the high-$\pt$ radiation
in {\sc Powheg}, is set to the top quark mass.
The CT10 \cite{Lai:2010vv} PDF are employed and the top quark mass is set to $m_\mathrm{top} = 172.5$ \GeV. 
Parton showering and hadronization are simulated with {\sc Pythia} v6.425
\cite{SAMPLES-PYTHIA} using the Perugia 2011 C set of tuned parameters (tune) \cite{Skands} 
and the corresponding leading-order (LO) CTEQ6L1 \cite{cteq6l} PDF set. 
Unless otherwise noted, electroweak corrections extracted with
{\sc Hathor 2.1-alpha}~\cite{Aliev:2010zk}, implementing the theoretical
calculations of Refs.~\cite{Kuhn:2005it,Kuhn:2006vh,Kuhn:2013zoa}, are
applied as weights to the events of this sample.
The prediction of {\sc Powheg} is compared to that obtained with
other generators such as
{\sc MC@NLO} v4.01 \cite{Frixione:2008ym} with CT10 for the PDF set, interfaced to {\sc Herwig} v6.520 \cite{SAMPLES-HERWIG} 
for parton showering and hadronization, {\sc Jimmy} v4.31 \cite{SAMPLES-JIMMY} for the modeling of multiple 
parton scattering. 
In Herwig and Jimmy the CT10 PDF is used and the ATLAS AUET2 tune \cite{ATL-PHYS-PUB-2011-008} is employed for the parton shower and hadronization settings.
In addition, the LO multileg generator
{\sc Alpgen} v2.13 \cite{SAMPLES-ALPGEN} interfaced to {\sc Herwig} is used
where up to four additional partons in the 
matrix element are produced;  
the MLM \cite{Mangano:2006rw} matching scheme is employed 
to avoid double counting of configurations generated by both the
parton shower and the matrix-element calculation; the  CTEQ6L1 \cite{cteq6l}
PDF set is employed; heavy-flavor quarks
are included in the matrix-element calculations to produce the $\ttbar + b\bar{b}$
and  $\ttbar + c\bar{c}$ processes; the overlap between the heavy-flavor quarks
produced from the matrix-element calculations and from the parton shower is removed.
For the evaluation of systematic uncertainties due to the parton
showering and hadronization models, 
a {\sc Powheg}+{\sc Herwig} sample is compared to a {\sc Powheg}+{\sc
  Pythia} sample. 
The uncertainties due to QCD initial- and final-state radiation (ISR
and FSR) modeling are estimated 
with samples generated with {\sc AcerMC} v3.8 \cite{Kersevan:2004yg}, interfaced to {\sc
  Pythia} for which the parton shower parameters are varied according
to a measurement of the additional jet activity in $\ttbar$ events \cite{ttveto}.
The tunes for samples used to describe $\ttbar$ production show a reasonable 
agreement over a broad range of observables and kinematic regions in $\ttbar$ events 
\cite{Aad:2013fba,ATL-PHYS-PUB-2013-005,ATL-PHYS-PUB-2015-002}.
The electroweak corrections that are applied to the nominal {\sc Powheg+Pythia} sample are not 
applied to the other samples. The $\ttbar$ samples are normalized to the
NNLO+NNLL cross-section\footnote{
  The top++2.0 \cite{Czakon:2011xx} 
  calculation includes the next-to-next-to-leading-order QCD
  corrections and resums next-to-leading logarithmic soft gluon
  terms. The quoted cross-section corresponds to a top quark mass of
  172.5 \GeV.} 
\cite{Cacciari:2011hy,Beneke:2011mq,Baernreuther:2012ws,Czakon:2012zr,Czakon:2012pz,Czakon:2013goa}: 
$\sigma_{t\bar{t}}=253^{+13}_{-15}$~pb.

Leptonic decays of vector bosons produced in association with several high-$\pt$ jets,
referred to as $W$+jets and $Z$+jets, constitute the largest
background in this analysis. 
Samples of simulated $W/Z$+jets events with up to five additional
partons in the LO 
matrix elements are produced with the {\sc Alpgen} generator 
interfaced to {\sc Pythia} for parton showering using the MLM matching
scheme. Heavy-flavor quarks
are included in the matrix-element calculations to produce the $Wb\bar{b}$,
$Wc\bar{c}$, $Wc$, $Zb\bar{b}$, and $Zc\bar{c}$ processes.
 The overlap between the heavy-flavor quarks
produced by the matrix element and by parton showering is removed.
$W$+jets samples are normalized 
to the inclusive $W$ boson NNLO cross-section \cite{Hamberg:1990np,Gavin:2012sy} and corrected by 
applying additional scale factors 
derived from data, as described in Sec.~\ref{sec:bkgd}.
 
Single top quark production in the $t$-channel is simulated using the {\sc AcerMC} generator, 
while production in the $s$-channel and the production of a top quark in association with
a $W$ boson are modeled with 
{\sc Powheg} \cite{Alioli:2009je,Frederix:2012dh,Re:2010bp,Frixione:2008yi}.
Both generators are interfaced with {\sc Pythia} using the CTEQ6L1 PDF set 
and the Perugia 2011 tune for parton shower modeling. 
The cross-sections multiplied by the branching ratios for the leptonic $W$ decay employed for these processes are 28.4 pb ($t$-channel) \cite{Kidonakis:2011wy},
22.4 pb ($Wt$ production) \cite{Kidonakis:2010ux}, and 1.8 pb ($s$-channel) \cite{Kidonakis:2010tc}, 
as obtained from NLO+NNLL calculations.

Diboson production is modeled using {\sc
  Sherpa}~\cite{Gleisberg:2008ta} with the {\sc CT10} PDF set 
and the yields are normalized to the NLO cross-sections \cite{Campbell:2011bn}.

%% file: objdefinitions.tex
\section{Object definition and event selection}
\label{sec:objdef}

Jets are reconstructed using the anti-$k_t$ algorithm \cite{Cacciari:2008gp} 
implemented in the {\sc FastJet} package
\cite{Cacciari:2011ma} with radius 
parameter $R = 0.4$ or $R = 1.0$, respectively called small-$R$ and large-$R$ jets in the following,
using as input calibrated topological
clusters \cite{topocluster, Aad:2011he, CONF-2015-017}. 
 These clusters are assumed to be massless when computing the
 jet four-vectors and substructure variables.
Large-$R$ jets containing  hadronically decaying top quarks are
selected by applying jet substructure requirements, which exploit the fact
that they contain several high-$\pt$ objects and have a high mass, unlike
most jets originating from the fragmentation of other quarks or
gluons. 
The trimming algorithm \cite{Krohn:2009th} 
with parameters $R_\mathrm{sub} = 0.3$ and $f_\mathrm{cut} = 0.05$ is applied to large-$R$ jets 
to mitigate the impact of  initial-state radiation, underlying-event activity, and pileup.
A correction
for the 
number of additional $pp$ interactions per bunch crossing is
applied to small-$R$ jets
\cite{Cacciari:2008gn,Cacciari:2007fd,Ellis:1993tq,Catani:1993hr}.
The $\pt$ of small-$R$ jets and large-$R$ trimmed jets and the large-$R$ jet mass, obtained from the four-momentum sum
of all jet constituents, are calibrated using energy- and
$\eta$-dependent correction factors. 
After this calibration, the $\pt$ and mass of the jets in simulated events correspond on average to the ones of the corresponding particle-level jets,
which are built from the stable particles produced by the MC event generator  \cite{Aad:2014bia,Aad:2013gja}.
Differences between the small-$R$ jet response
in data and MC simulation are evaluated from control samples 
and corresponding corrections are applied to data.
Small-$R$ jets are required to be in the fiducial region $| \eta | <
2.5$ and must have $\pt > 25$ \GeV.
The jet vertex fraction (JVF) is a measure of the fraction of the
jet's track momenta that originate from the primary vertex. It is computed as the summed $\pt$ of all tracks 
matched to the jet and the primary vertex, divided by the summed $\pt$ of all tracks 
matched to the jet. Small-$R$ jets with $\pt < 50$ \GeV\ and $|\eta| <
2.4$ are rejected when $\mathrm{JVF} < 0.5$, to reduce the contribution 
of jets generated by pileup interactions.\footnote{The jet is
  retained if no tracks are assigned
to the jet.}
Trimmed large-$R$ jets are considered for the analysis if $| \eta | <
2.0$ and $\pt > 300$~\GeV. More details on the reconstruction and
performance of highly boosted top quarks in ATLAS can be found in
Ref.~\cite{Aad:2013gja,ATLAS-CONF-2015-036}.  

Small-$R$ jets containing a $b$-hadron are tagged using a 
neural-network-based algorithm (MV1) \cite{CON-2014-046} that combines information from
the track impact parameters, secondary vertex location, and 
decay topology inside the jets.
The operating point corresponds to an overall 70\% $b$-tagging
efficiency in $\ttbar$ events, and to a 
probability to mistag light-flavor jets of approximately 1\%. 

Electron candidates are reconstructed as charged-particle tracks in 
the inner detector associated with energy deposits in the EM
calorimeter. They must satisfy identification criteria based on the
shower shape in the EM calorimeter,  
on track quality, and on the transition radiation observed in the TRT detector~\cite{Aad:2014fxa}.
Electrons are required to be in the pseudorapidity region $| \eta | < 2.47$, 
excluding the transition region between the barrel and the endcap calorimeters 
($1.37 < | \eta | < 1.52$). The EM clusters must have a transverse energy $E_\mathrm{T} > 25$ \GeV.
The associated track must have a longitudinal impact parameter $|z_0|<$~2~mm 
with respect to the primary vertex, which is the vertex with the
highest $\sum \pt^2$ of the associated tracks in the event.

Muon candidates are defined by matching track segments in the muon spectrometer 
with tracks in the inner detector. The track $\pt$ is determined through a global fit of 
the track that takes into account the energy loss in the calorimeters~\cite{Aad:2014rra}.
The track is required to have a longitudinal impact parameter $|z_0|<$ 2 mm,
and a transverse impact parameter significance $|d_0/\sigma(d_0)|~<~3$, indicating the track is consistent with originating from the hard-scattering vertex.
Muons are required to have $\pt > 25$ \GeV\ and be in the fiducial region $| \eta | < 2.5$.

Lepton candidates are required to be isolated to suppress  background leptons
originating from jets.  
The variable ``mini-isolation'' \cite{Rehermann:2010vq} is used. It is defined as $I_\mathrm{mini} = \sum_{\mathrm{tracks}} \pt^{\mathrm{track}} / \pt^\ell$, where 
$\pt^\ell$ is the lepton transverse momentum and the sum is over all
good-quality tracks (excluding the lepton track) that 
have  $\pt > 0.4$ \GeV\ and a distance from the lepton $\Delta
R = \sqrt{(\Delta\eta)^2 + (\Delta\phi)^2}  < K_\mathrm{T} / \pt^\ell$.
The parameter $K_\mathrm{T}$ is set to 10 \GeV\ and the isolation requirement $I_\mathrm{mini} < 0.05$ is applied for both the 
electrons and muons. An isolation cone that decreases in size with
increasing $\pt^\ell$ improves the selection efficiency of the 
decay of high-$\pt$ top
quarks.

Since leptons deposit energy in the calorimeters, an overlap removal procedure is applied 
in order to avoid double counting of leptons and small-$R$ jets. 
In order to improve the reconstruction efficiency in the highly boosted topology, the same overlap removal procedure as used in Ref.~\cite{Aad:2015fna} has been adopted.
First, jets close to electrons, with $\Delta R(e,\mathrm{jet}_{R=0.4}) < 0.4$ are corrected 
by subtracting the electron four-vector from the jet four-vector and the JVF is recalculated after removing 
the electron track. The new $e$-subtracted jet is retained if it satisfies the jet selection criteria 
listed above, otherwise it is rejected. After this procedure, electrons that lie 
within $\Delta R(e,\mathrm{jet}_{R=0.4}) = 0.2$ from a small-$R$ jet
are removed and their four-momentum added back to that of the jet. 
The muon--jet overlap removal procedure removes muons that fall inside a 
cone of size $\Delta R(\mu,\mathrm{jet}_{R=0.4}) < 0.04 +10\mathrm{~\GeV}/p_{\mathrm{T},\mu}$
around a small-$R$ jet axis. 

The missing transverse momentum $\met$ is the magnitude of the vector 
sum of the transverse energy of all calorimeter cells \cite{Aad:2012re}. Their energy is corrected 
on the basis of the associated physics object.
The contribution of muons is added using their transverse momentum obtained from the 
tracking system and the muon spectrometer.

The event selection proceeds as follows.
Each event
must have a reconstructed primary vertex with five or more associated
tracks with $\pt > 0.4$~\GeV.
The events are required to contain exactly one reconstructed
lepton candidate with $\pt > 25 \gev$. 
The transverse mass of the lepton and 
$\met$ is defined as $m_\mathrm{T}^W = \sqrt{ 2 \pt^\ell \met
 (1 - \cos \Delta\phi) }$, where  $\Delta\phi$ is the azimuthal angle
between the lepton and $\met$. 
Events are retained if $\met > 20$ \GeV\ and $\met~+~m_\mathrm{T}^W~>~60 \gev$ to
suppress QCD multijet events.

 The selection exploits the fact that the highly boosted top quark decay
 products tend to be collimated.
Therefore events are selected by requiring the presence of at least one small-$R$ 
jet close to the lepton ($\Delta R(\ell,\mathrm{jet}_{R=0.4}) < 1.5$) and the existence of a reconstructed 
large-$R$ trimmed jet with mass $m_\mathrm{jet} > 100$~\GeV. 
To improve the rejection of background jets, originating from light quarks or gluons, a cut on 
the $k_t$ splitting scale \cite{Ellis:1993tq,Catani:1993hr} of the large-$R$ jets is made. 
The $k_t$ splitting scale is calculated by reclustering the large-$R$ jet with the $k_t$-clustering 
algorithm, and taking the $k_t$ distance between the two subjets of the final clustering step to be 
$\sqrt{d_{12}} = \min (p_{\mathrm{T}1},p_{\mathrm{T}2}) \Delta
R_{12}$, where $p_{\mathrm{T}1}$ and $p_{\mathrm{T}2}$ are the
transverse momenta
of the two subjets and $\Delta R_{12}$ is the distance between them. 
It is expected to have large values for jets containing two hard subjets, as expected in the decay of massive objects.
Events are selected if the large-$R$ jet has $\sqrt{d_{12}} > 40$~\GeV.
The large-$R$ jet must be well separated from the lepton ($\Delta \phi(\ell,\mathrm{jet}_{R=1.0}) > 2.3$) 
and from the small-$R$ jet associated with the lepton ($\Delta
R(\mathrm{jet}_{R=1.0},\mathrm{jet}_{R=0.4}) > 1.5$). 
The leading-$\pt$ trimmed large-$R$ jet satisfying these requirements
is referred to as the {\it top-jet} candidate. 
Finally, at least one of the two top quark candidates must be $b$-tagged.
This implies that either 
the highest-$\pt$ small-$R$ jet close to the lepton ($\Delta R(\ell,\mathrm{jet}_{R=0.4}) < 1.5$)
or at least one small-$R$ jet close to the large-$R$ jet ($\Delta R(\mathrm{jet}_{R=1.0},\mathrm{jet}_{R=0.4}) < 1.0$) 
is $b$-tagged\footnote{The reconstruction of a large-$R$ jet does not
prevent the reconstruction of small-$R$ jets overlapping with it.}. 

The event selection is summarized in Table 1.
After these requirements the data sample contains 4145 and 3603 events
in the electron channel and muon channel, respectively, of which $\approx
85\%$ are expected to be semileptonic
$\ttbar$ events.

\begin{table*} [!htbp]
\noindent\makebox[\textwidth]{
  \begin{tabular}{|l|l|l|l|}
  \hline
  Cut & \multicolumn{2}{|c|}{Detector level} & Particle level \\
  \hline
            & $\eplus$ & $\muplus$ & \\
  \hline\hline
  Leptons & \parbox[c][5em][c]{5cm}{$|z_0|<$ 2 mm \\$I_\mathrm{mini}
    <$ 0.05\\ $|\eta|<$1.37 or 1.52$<|\eta|<$2.47 \\ $p_{\mathrm{T}} >
    $ 25 \GeV} & \parbox[c][7em][c]{5.5cm}{$|z_0|<$ 2 mm and $|d_0/\sigma(d_0)|<$ 3 \\$I_\mathrm{mini} <$ 0.05\\ $|\eta|<$2.5 \\$p_{\mathrm{T}} > $ 25 \GeV } & \parbox[c][7em][c]{2.0cm}{$|\eta|<$2.5\\ $p_{\mathrm{T}} > $ 25 \GeV} \\
  \hline
  Anti-$k_t$ $R=0.4$ jets &
  \multicolumn{2}{|c|}{\parbox[c][4em][c]{6.5cm}{\centering  $p_{\mathrm{T}} > $ 25
      \GeV \\  $|\eta|<$2.5 \\ $\mathrm{JVF} > 0.5$ (if $p_{\mathrm{T}}<$ 50 \GeV\ and $|\eta| <
2.4$)}} & \parbox[c][4em][c]{2.0cm}{$|\eta|<$2.5\\ $p_{\mathrm{T}} > $ 25 \GeV} \\
  \hline
  Overlap removal & \parbox[c][7em][c]{5.5cm}{if $\Delta
    R(e,\mathrm{jet}_{R=0.4})<$0.4: \\ $\mathrm{jet}_{R=0.4}' =
    \mathrm{jet}_{R=0.4} - e$\\ if $\Delta
    R(e,\mathrm{jet}_{R=0.4}')<$0.2:\\ $e$ removed
   and
  \\ $ \mathrm{jet}_{R=0.4}'' =\mathrm{jet}_{R=0.4}' + e$ }& \parbox[c][4em][c]{7.0cm}{if $\Delta
    R(\mu,\mathrm{jet}_{R=0.4}')<0.04 +
    10\mathrm{~\GeV}/p_{\mathrm{T}}(\mu)$: \\$\mu$ removed  }& None \\
  \hline 
  \met,\mtw & \multicolumn{3}{|c|}{ \met $>$ 20 \GeV, \met+\mtw $>$ 60 \GeV}   \\
  \hline
  Leptonic top & \multicolumn{3}{|c|}{ At least one anti-$k_t$ $R = 0.4$ jet with $\Delta R(\ell,\mathrm{jet}_{R=0.4}) < 1.5$ }\\ 
  \hline
  Hadronic top &
  \multicolumn{3}{|c|}{ \parbox[c][4em][c]{13cm}{\centering The
      leading-$\pt$ trimmed anti-$k_t$ $R = 1.0$ jet has: \\$p_{\mathrm{T}}>$ 300 \GeV, $m>$ 100 \GeV, $\sqrt{d_{12}}>$ 40 \GeV \\ $\Delta R(\mathrm{jet}_{R=1.0},\mathrm{jet}_{R=0.4}) > 1.5$, $\Delta \phi(\ell,\mathrm{jet}_{R=1.0}) > 2.3$}} \\
  \hline
  $b$-tagging &
  \multicolumn{3}{|c|}{ \parbox[c][4em][c]{13cm}{\centering At least
      one of: \\1) the leading-$\pt$ anti-$k_t$ $R = 0.4$ jet with $\Delta R(\ell,\mathrm{jet}_{R=0.4}) < 1.5$ is $b$-tagged \\  2) at least one anti-$k_t$ $R = 0.4$ jet with $\Delta R(\mathrm{jet}_{R=1.0},\mathrm{jet}_{R=0.4}) < 1.0$ is $b$-tagged}} \\
  \hline
  \end{tabular}
}
\label{tab:seln}
\caption{Summary of event selections for detector-level and MC-generated
  particle-level events described in Secs.~\ref{sec:objdef} and
  \ref{sec:fiducialregion}, respectively.}
\end{table*}

%% file: backgrounds.tex
\section{Background estimation}
\label{sec:bkgd}

After the event selection the background is
composed primarily, in order of importance, 
of $W$+jets, $\ttbar$~dilepton, single top, and QCD multijet events. The $W$+jets
background is obtained from MC simulation with normalization and
heavy-flavor content adjusted in data control regions. The $\ttbar$
dilepton background is determined as a fraction of the full $\ttbar$  sample
predicted by MC simulation.  
QCD multijet events are estimated with a fully data-driven
method. Single top production as well as minor backgrounds ($Z$+jets
and diboson) are determined from MC simulation normalized to the best
available theoretical calculation of their cross-sections.

The $W$+jets background estimate uses as a
starting point the {\sc Alpgen+Pythia}  samples normalized to
the inclusive $W$ boson NNLO cross-section. 
The normalization and heavy-flavor fraction of the $W$+jets background
have large theoretical uncertainties, and are then 
determined from data. 
The overall $W$+jets normalization is obtained by exploiting the
expected charge asymmetry in the production of $W^+$ and $W^-$ bosons at a
$pp$ collider \cite{Aad:2012hg,ATLAS:2012an}. This asymmetry is predicted precisely by theory, and other
processes in the $\ttbar$ sample are symmetric in
charge except for a small contamination from
single top and $WZ$ events, which is corrected by MC simulation. The total number of $W$+jets events in the sample can thus be
estimated with the following equation:
\begin{equation}
N_{W^+} + N_{W^-} = \left(\frac{r_\mathrm{MC} + 1}{r_\mathrm{MC} - 1}\right)(D_\mathrm{+} - D_\mathrm{-}),
\label{eq:Wchargeasymm}
\end{equation}
where $r_\mathrm{MC}$ is the ratio of the number of events with positive leptons to
the number with negative leptons in the MC simulation, and $D_\mathrm{+}$ and $D_\mathrm{-}$ are the
number of events with positive and negative leptons in the data,
respectively. The signal sample has too few events 
 to apply Eq.~(\ref{eq:Wchargeasymm}) 
directly. Instead a sample enhanced in $W$+jets events is obtained by
removing the $b$-tagging, $\Delta \phi(\mathrm{jet}_{R=1.0},\ell)$, jet mass, and
$\sqrt{d_{12}}$ requirements.
The heavy-flavor fraction scale factors correct for potential mismodeling in the generator 
of the fractions of $W$ production associated with different flavor components ($W+b\bar{b}$, $W+c\bar{c}$, $W+c$). 
They are estimated in a sample with the same
lepton and $\met$ selections as the signal selection, but with only
two small-$R$ jets and no
$b$-tagging requirements. The $b$-jet multiplicity, in conjunction
with knowledge of the $b$-tagging and mistag efficiency, is used
to extract the heavy-flavor fraction in this sample. A common scale
factor is used for the $W+b\bar{b}$ and $W+c\bar{c}$ components.  This information
is extrapolated to the signal region using the MC simulation, assuming
constant relative rates for the signal and control regions. 
The overall normalization and heavy-flavor scale factors are extracted
iteratively because the various flavor components have different charge asymmetries.
After correction the $W$+jets events are expected to make up approximately
5\% of the total events in the signal region. 

QCD multijet events can mimic the 
lepton+jets signature. This background is estimated directly
from data by using the matrix-method technique \cite{ATLAS-CONF-2014-058}.
A sample enhanced in fake leptons, i.e., nonprompt leptons or 
jets misidentified as prompt leptons, is obtained by loosening the lepton
identification requirements. 
The number of events with fake leptons in the signal region can be
predicted as:
\begin{linenomath}
 \begin{equation*}
 N_\mathrm{multijet}=\frac{(\epsilon-1)\;f}{\epsilon-f}N_\mathrm{T}+\frac{\epsilon\;f}{\epsilon-f}N_\mathrm{L},
 \label{eqn:QCDMM}
 \end{equation*}
\end{linenomath}
where $\epsilon$ and $f$ are the efficiencies for
leptons that passed the loose selections to also pass the tight (signal)
selections, for real and fake leptons respectively, $N_\mathrm{T}$ is the number of events with a tight
lepton, and $N_\mathrm{L}$ is the number of events with a loose lepton that failed the tight cuts. The efficiency $f$ is
measured using data in fake-lepton-enhanced control regions and $\epsilon$ is extracted
from MC simulation and validated in data.
QCD multijet events contribute to the total event yield at
approximately the
percent level.

Top quark pair events with both the top and antitop quarks decaying leptonically (including decays to $\tau$) can sometimes pass the 
event selection, contributing approximately 5\% of the total event
yield, and are treated as background in the analysis. The fraction of dileptonic $\ttbar$ events
in each $\pt$ bin is estimated using the same MC sample used to model
the signal.

%% file: systematics.tex
\section{Systematic uncertainties}
\label{sec:syst}

Systematic uncertainties, which arise from object reconstruction and
calibration, 
MC generator modeling, and background estimation, are described
below. The propagation of systematic uncertainties
through the unfolding procedure is described in Sec.~\ref{sec:subsyst}.

\subsection{Detector modeling}
\label{sec:detSyst}

The uncertainty on the large-$R$ jet energy scale (JES), jet mass
scale (JMS), and $k_t$ splitting scale
is
obtained using two different data-driven methods. 
For $\pt > 800$ \GeV\ for JES, and for all $\pt$ for the JMS and $k_t$ splitting scale, 
the ratio of the large-$R$ jets kinematic variables reconstructed from the calorimeter 
clusters to those from inner-detector tracks is compared between data and MC simulation.
For $\pt < 800$ \GeV\ for JES, the $\pt$ of large-$R$ jets are 
compared to the well-calibrated $\pt$ of photons in a large
sample of photon+jets events. 
An additional MC-based uncertainty, referred to as large-$R$ JES topology
uncertainty, is included to reflect the fact that the jets in these
calibration samples have a different response (gluon or light-quark
jets) than those in $\ttbar$ events (top-jets). The full difference
between the response to these two types of jets is conservatively
assigned as the corresponding systematic uncertainty.
The uncertainty on the large-$R$ jet energy resolution (JER) is determined by smearing the
jet energy such that the resolution is degraded by 20\%
\cite{ATLAS:2012am,Aad:2012ag} and evaluating 
the effect on the final result.  The same smearing procedure is
applied to determine the uncertainty due to the large-$R$ jet mass resolution (JMR). 
The uncertainties on the large-$R$ jets JES are the dominant contribution to the total
uncertainty of this measurement, in particular the topology and photon+jet
calibration uncertainties.

The small-$R$ jet energy scale uncertainty is derived using a combination of simulations,
test beam data, and {\it in situ} measurements \cite{Aad:2014bia,Aad:2011he,Aad:2012vm}. Additional contributions
from the jet flavor composition, calorimeter response to different jet flavors, 
and pileup are taken into
account. Uncertainties in the jet energy resolution are
obtained with an {\it in situ} measurement of the jet $\pt$ balance 
in dijet events \cite{JER}.

The efficiency to tag $b$-jets and mistag light jets is corrected in 
Monte Carlo events by applying $b$-tagging scale factors, extracted in 
$\ttbar$ and dijet samples,
that compensate for the residual difference between data and simulation.
The associated systematic uncertainty is computed by varying the scale factors 
within their uncertainty \cite{ATLAS-CONF-2014-004,CON-2012-043, CONF-2012-040}. 
The $b$-jet calibration is performed for jets with $\pt$ up to 300 \GeV; for larger transverse
momenta an additional MC-based extrapolation uncertainty is applied, which ranges
from approximately 10\% to 30\%, increasing with $b$-jet 
$\pt$ from 300 \GeV\  to 1200 \GeV. 

The lepton reconstruction efficiency in simulation is corrected by scale factors 
derived from measurements of these efficiencies in data using $Z \to \ell^+ \ell^-$ 
enriched control regions. 
The lepton trigger and reconstruction efficiency scale factors, energy scale, and energy resolution are varied 
within their uncertainties \cite{Aad:2011mk,Aad:2014rra}.

The uncertainty associated with $\met$ is calculated by propagating the 
energy scale and resolution systematic uncertainties on all physics objects to the 
$\met$ calculation. Additional $\met$ uncertainties arising from energy deposits not associated with any reconstructed objects are also included \cite{Aad:2012re}.

The uncertainty on the integrated luminosity is $\pm$ 2.8\% and is
derived following a methodology similar to that defined in
Ref.~\cite{Aad:2013ucp}.

\subsection{Signal and background modeling}
\label{sec:sigAndBkgSyst}

The $\ttbar$ parton shower and hadronization uncertainty is computed by comparing the results
obtained with {\sc Powheg+Pythia} (without electroweak corrections applied) and  {\sc Powheg+Herwig}. The
$\ttbar$ generator uncertainty is evaluated by taking the difference
between the results obtained with   {\sc Powheg+Herwig} and {\sc MC@NLO+Herwig}. Both
uncertainties are symmetrized. 
The procedure to compute the PDF uncertainty on the signal is based on the PDF4LHC
recommendations \cite{Botje:2011sn} using 
the {\sc MC@NLO+Herwig} sample
with three different PDF sets (CT10 \cite{Lai:2010vv}, MSTW \cite{Martin:2009iq} and NNPDF \cite{Ball:2012cx}).
An intra-PDF uncertainty is obtained for each PDF set by following its respective prescription while 
an inter-PDF uncertainty is computed as the envelope of the three intra-PDF uncertainties. 
The modeling of  ISR and FSR is evaluated separately using 
dedicated  {\sc AcerMC + Pythia} samples with variation of the 
{\sc Pythia} parameters for QCD radiation. 

The  $W$+jets shape uncertainty is
extracted by varying the renormalization and matching scales in {\sc Alpgen}.
The $W$+jets MC statistical uncertainty is also computed and 
its contribution to the cross-section uncertainty increases with the 
top-jet candidate $\pt$ from approximately 1\% to 6\%. 
A new set of $W$+jets normalization and heavy-flavor scale factors 
is extracted for each variation of the most important detector modeling
uncertainties, allowing their correlated
effect on the $W$+jets background, $\ttbar$ signal and background, and
other MC-based background processes to be assessed. 

The uncertainty on the fake-lepton background is determined by varying
the definition of loose leptons, changing the selection used to form
the fake-enhanced control region, and propagating the statistical uncertainty of
parameterizations of the efficiency and the fake rate. 

The single-top background is assigned an
uncertainty associated with the theory calculations used for its
normalization
\cite{Kidonakis:2011wy,Kidonakis:2010ux,Kidonakis:2010tc}.
A generator uncertainty is included for the $Wt$ channel, which provides the
largest single-top contribution, by taking the difference between 
the yields predicted by {\sc Powheg} and
{\sc MC@NLO}. An 
uncertainty on the interference between the $\ttbar$ and $Wt$
processes is also included.
A conservative uncertainty of 50\% is applied to the normalization of
the subdominant $Z$+jets and
diboson backgrounds.

%% file: datamccomp.tex
\section{Data and MC comparison at detector level}
\label{sec:datamccomp}

Table~\ref{tab:yield} gives the number of observed and expected events
for each process, where the systematic
uncertainties on the background estimates, objects' energy scales and reconstruction
efficiencies, and MC statistics are
taken into account. The prediction is generally found to overestimate
the data by approximately one standard deviation. 

\input{yieldstableNEW}

\begin{figure*}[htbp]
\centering
        \subfigure[]{\includegraphics[width=0.38\textwidth]{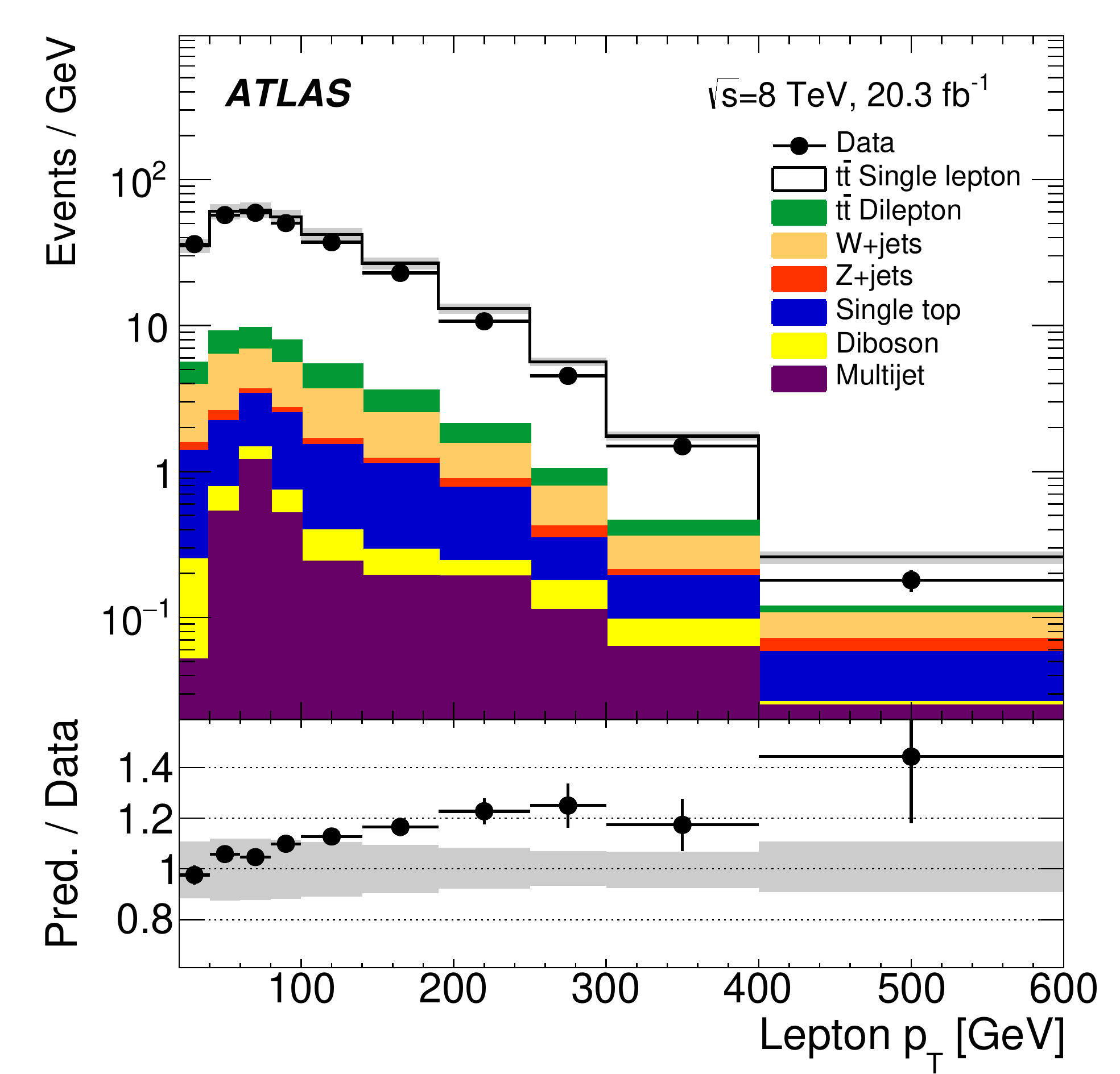}\label{h_lep_pt_EM}}
        \subfigure[]{\includegraphics[width=0.38\textwidth]{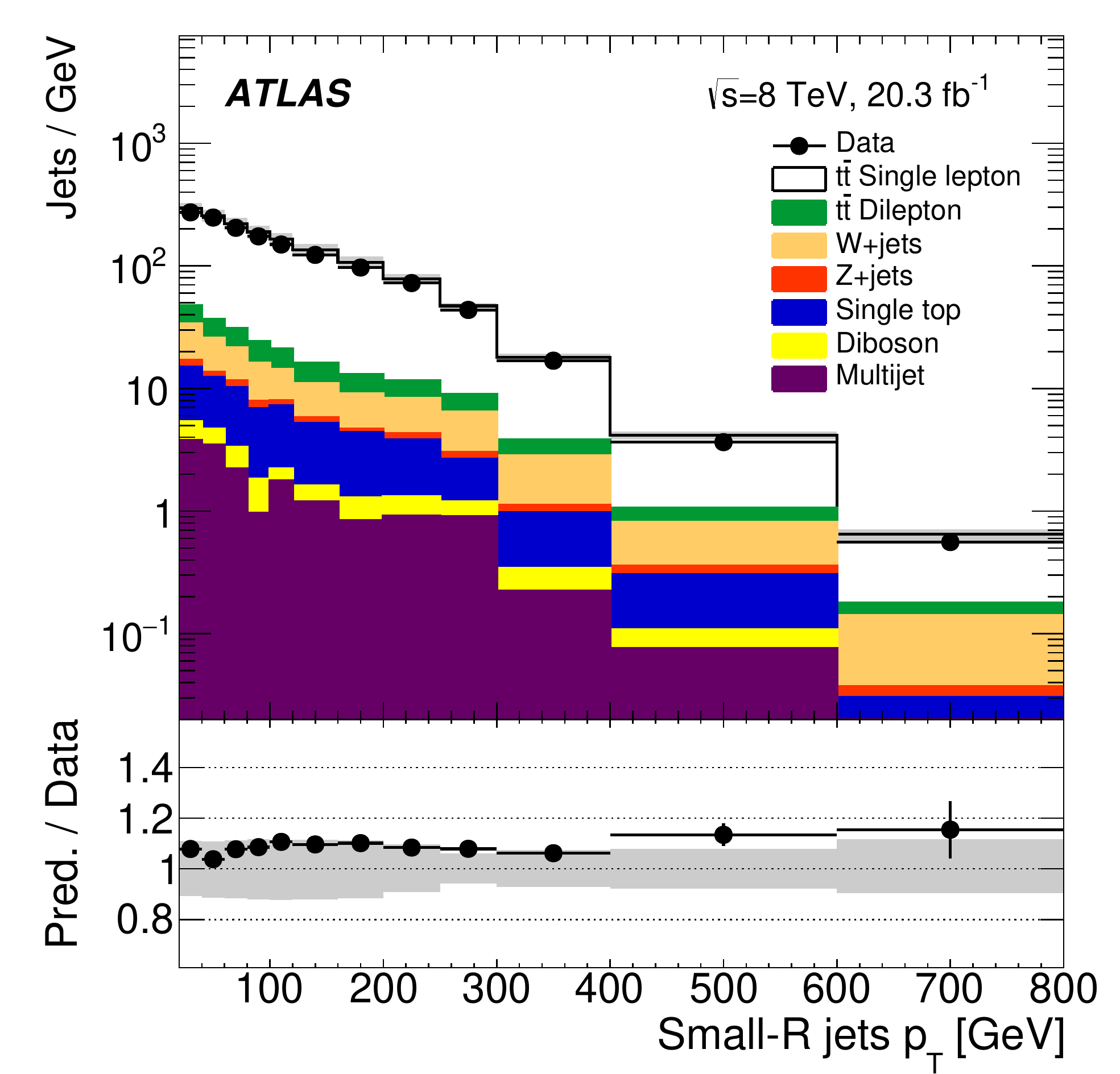}\label{h_akt4Jets_pt_EM}}\\
        \subfigure[]{\includegraphics[width=0.38\textwidth]{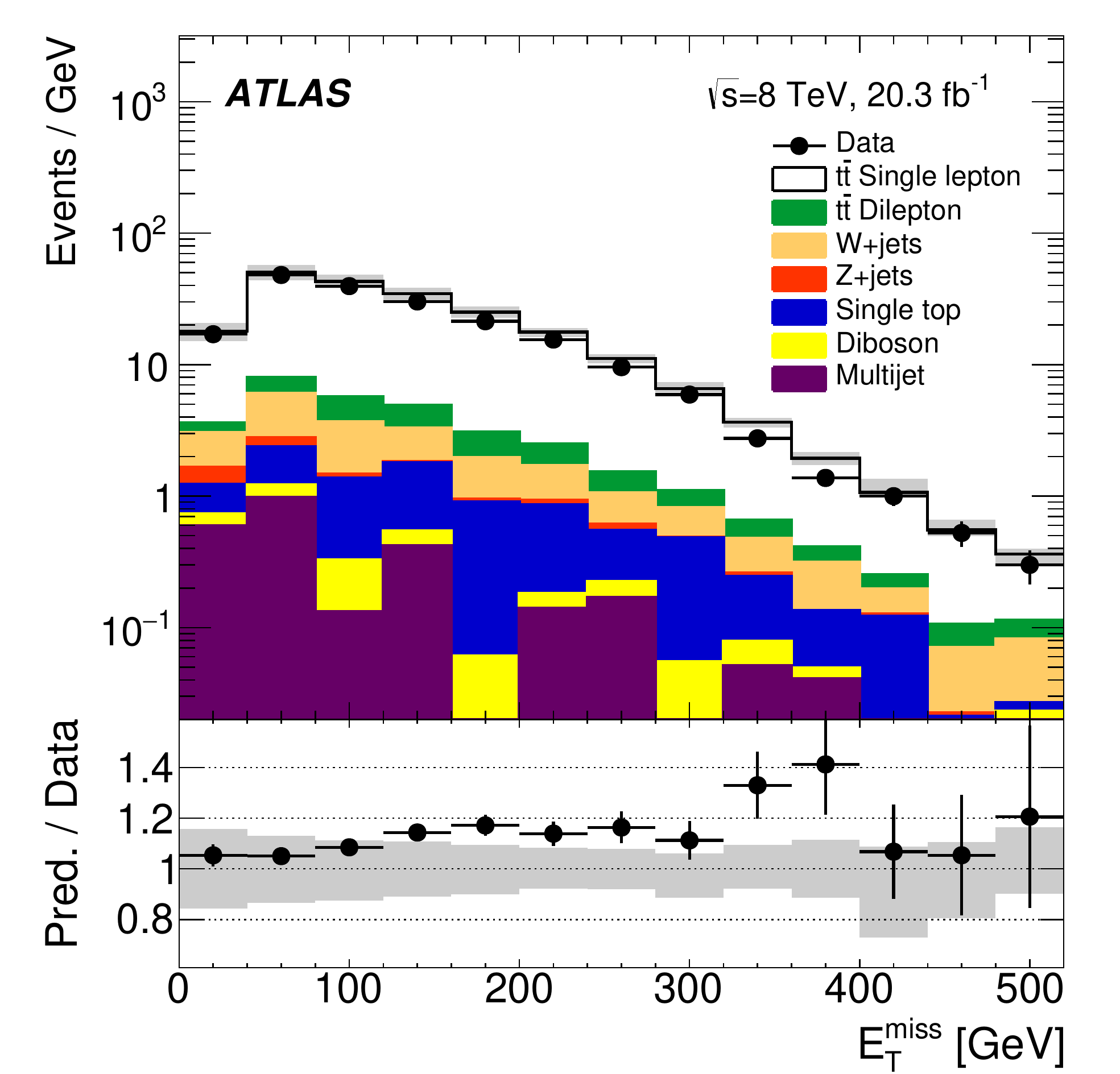}\label{h_met_EM}}
        \subfigure[]{\includegraphics[width=0.38\textwidth]{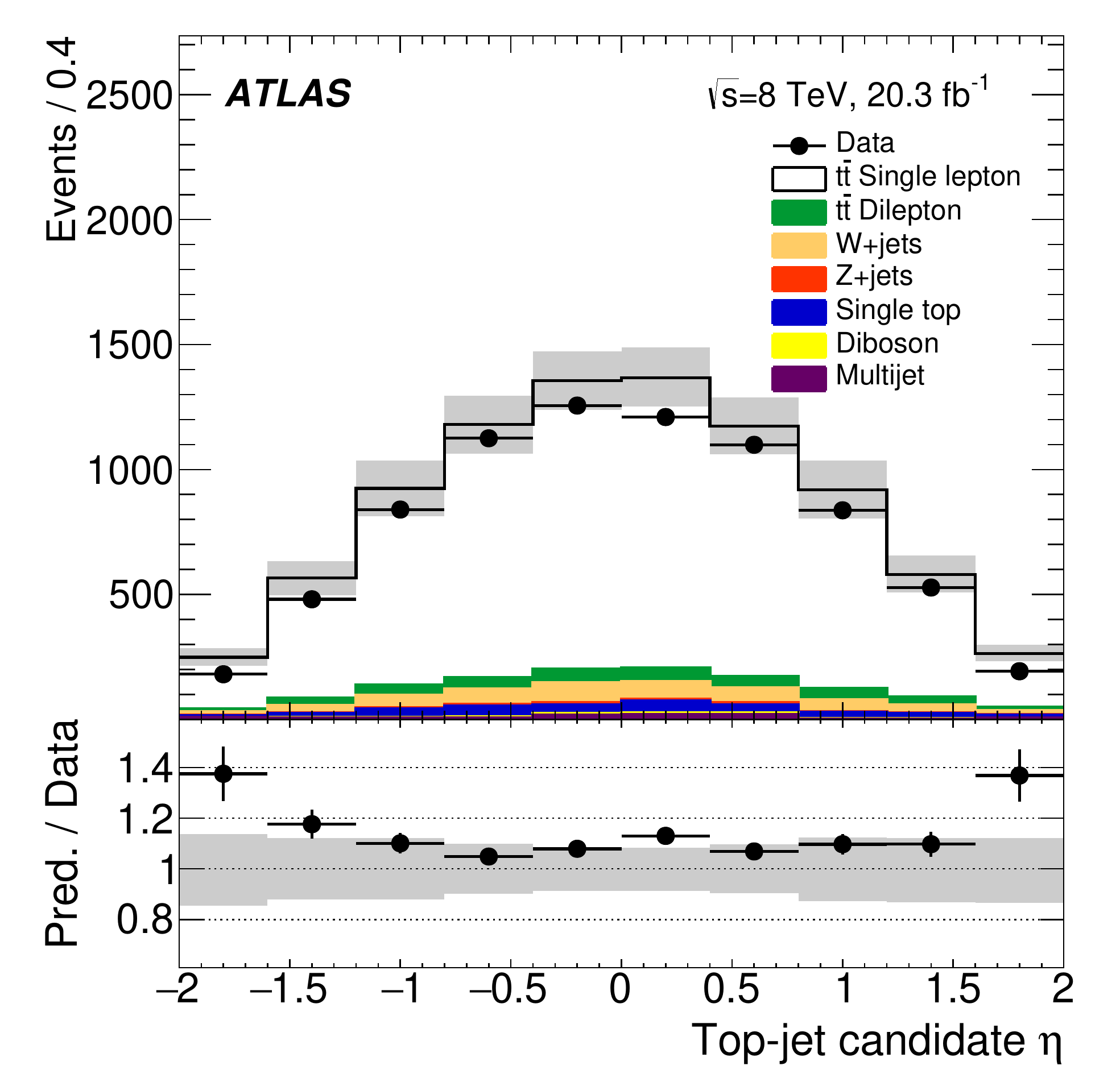}\label{h_recHadTop_eta_EM}}\\
        \subfigure[]{\includegraphics[width=0.38\textwidth]{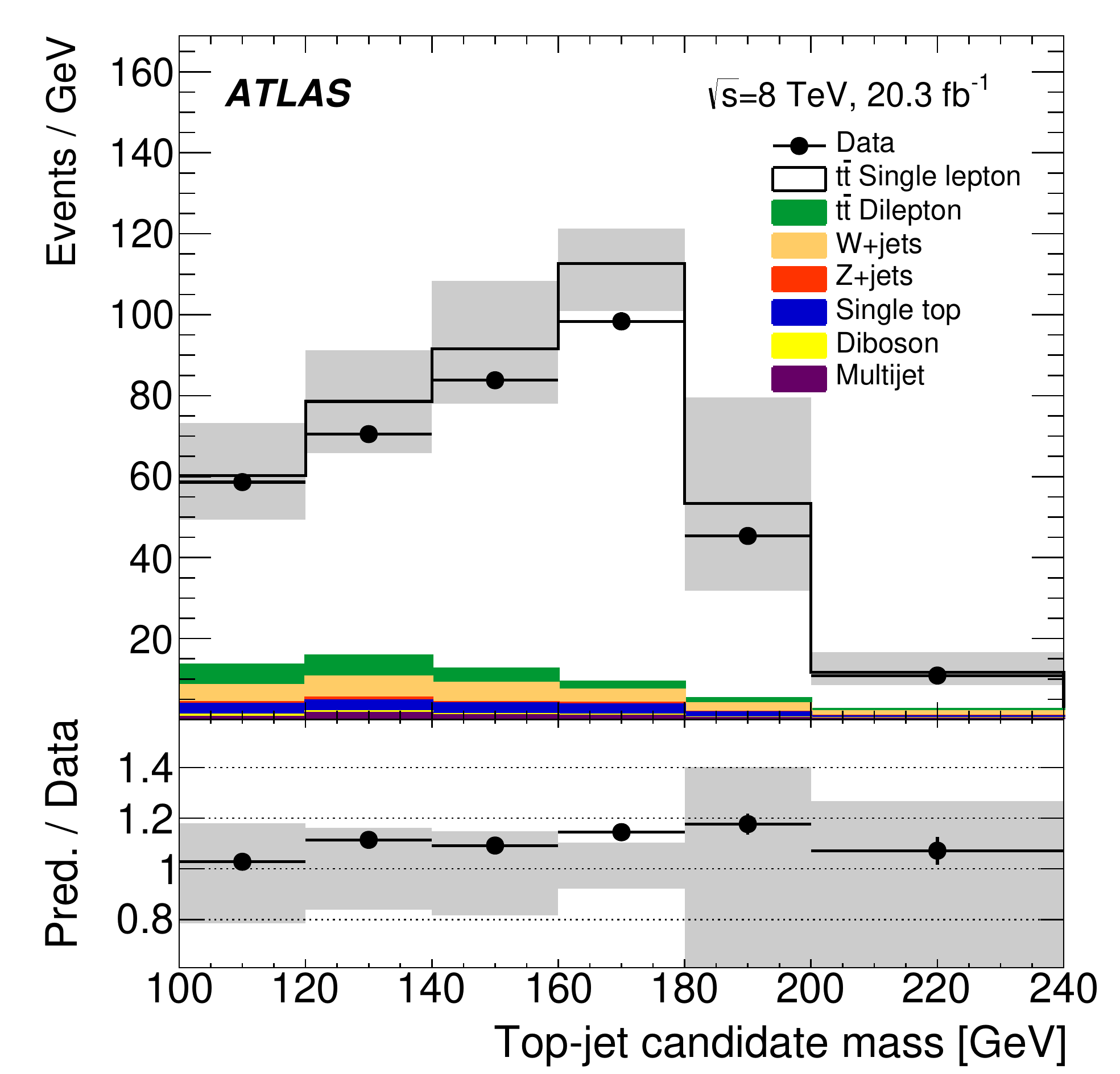}\label{h_recHadTop_m_EM}}
        \subfigure[]{\includegraphics[width=0.38\textwidth]{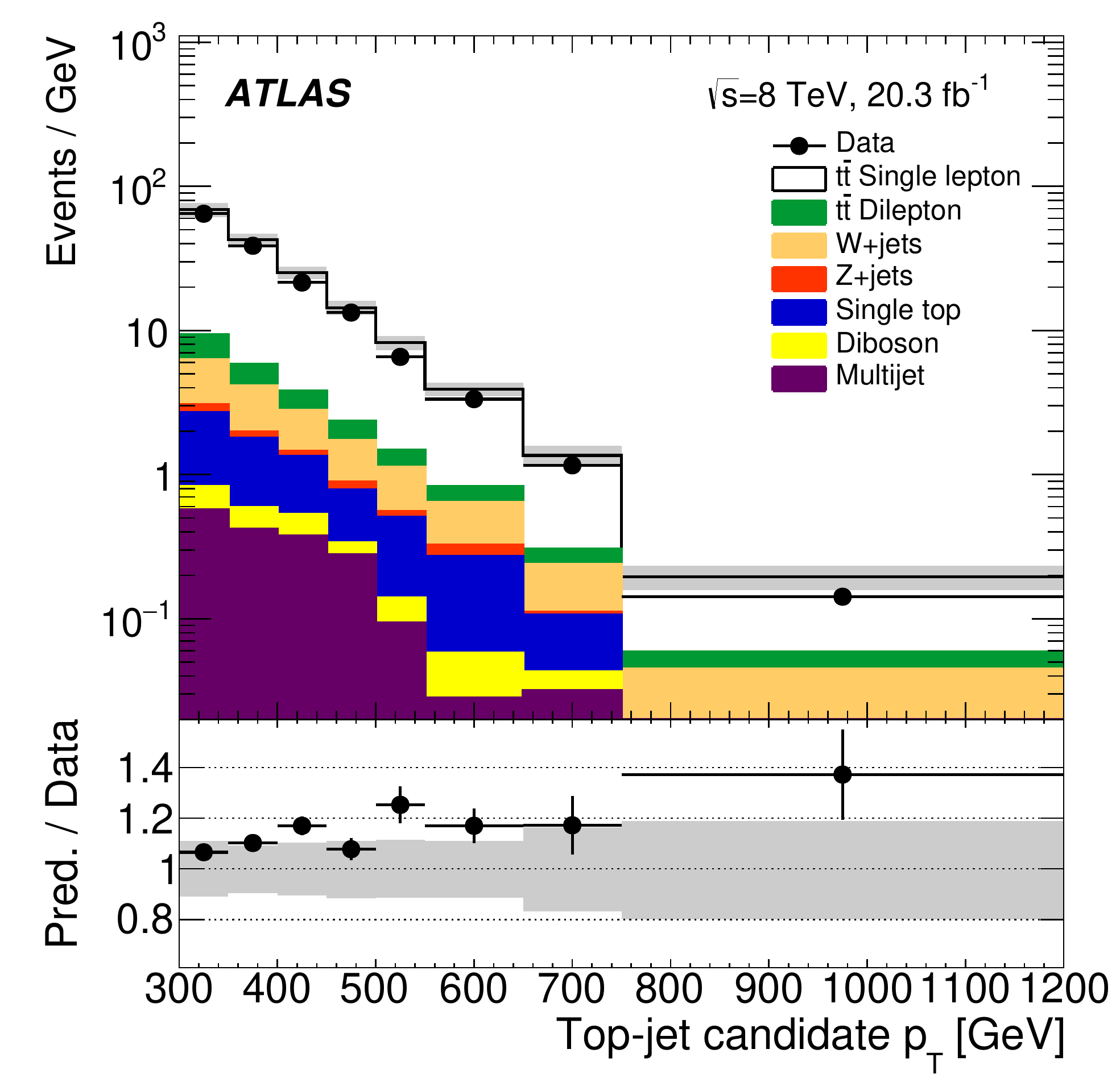}\label{h_recHadTop_pt_EM}}
       \caption{Distributions of (a) transverse momentum $\pt$ of the lepton candidates, (b) $\pt$ of
          selected small-$R$ jets, (c) missing transverse momentum $\met$,
	  (d) and pseudorapidity $\eta$, (e) mass and
          (f) $\pt$ of the leading selected anti-$k_t$ $R$=1.0 jets 
	  for the $\ell$+jets channel. The
          $\ttbar$ prediction is obtained using the nominal {\sc Powheg+Pythia} sample. The
	  ratio of the MC prediction to the data is shown in the insets below
          the histograms. The
          hashed area includes all the object-related uncertainties
          (on the jet, lepton, and $\met$), and the uncertainties from
          the background estimation, luminosity and MC statistics. The vertical lines indicate the data
          statistical uncertainty.}
        \label{fig:reco_all}
\end{figure*}

Agreement of the data with the prediction is further tested by studying the
distributions of several variables of interest in
Fig.~\ref{fig:reco_all}. The systematic uncertainties on the objects'
energy scales and reconstruction efficiencies, on the
background estimates, luminosity and MC statistics are shown. 
 While the prediction generally overestimates
 the data, as already seen in Table~\ref{tab:yield}, 
the simulation reproduces the observed shapes in most cases.
Exceptions include the tails of some kinematic variables such as the
 top-jet candidate $\pt$.
The distribution of the top-jet candidate $\pt$
constitutes the input to the unfolding procedure and is studied in
more detail in the following sections.

%% file: yieldstableNEW.tex
\begin{table}[tbh!]
\centering
\vspace{2mm}
\begin{tabular}{lcc}
\hline \hline
  & $e$+jets & $\mu$+jets \\
\hline
\ttb\ \ljets & 3880 $\pm$ 430    & 3420 $\pm$ 380\\
\ttb\ dilepton & 199 $\pm$ 27   & 169 $\pm$ 24\\
$W$+jets & 235 $\pm$ 54   & 226 $\pm$ 50\\
Single top & 133 $\pm$ 22   & 134 $\pm$ 29\\
Multijet & 91 $\pm$ 17   & 3 $\pm$ 1\\
$Z$+jets & 34 $\pm$ 18   & 14 $\pm$ 8\\
Dibosons & 22 $\pm$ 12   & 18 $\pm$ 10\\
\hline
Prediction & 4600 $\pm$ 470    & 3980 $\pm$ 410\\
Data & 4145  & 3603 \\
\hline\hline
\end{tabular}
\caption{Observed and expected number of events in the signal $e$+jets
  and $\mu$+jets samples. The systematic uncertainties include the
  background estimation techniques, objects' energy scales and reconstruction
efficiencies, and MC statistics.}
\label{tab:yield}
\end{table}

%% file: unfolding.tex
\section{Differential cross-section determination}
\label{sec:unfolding}

Differential cross-sections are measured as a function of the \pt\
of the top-jet candidate at particle level and the $\pt$  of the top
quark at parton level.  The electron and muon channels are first 
combined into a $\ell$+jets sample at the detector level. The
detector-level $\pt$ spectrum is corrected  for detector
inefficiencies and 
finite resolution to obtain 
particle-  and parton-level differential cross-sections. The particle-level measurement is performed in
a specific fiducial region of phase space close to the event
selection.  The  systematic and statistical
uncertainties are propagated through the unfolding
procedure. Finally a covariance matrix is computed to perform a 
quantitative comparison of the measured
cross-sections with MC predictions.

\subsection{Combination of decay channels}
\label{sec:combination}

The  $e$+jets
and $\mu$+jets selections are combined into a $\ell$+jets sample at the detector level.
The combined $\ell$+jets signal and background samples take into account the
efficiencies of the two selections. 
This procedure is well motivated given that the relative yields of the two channels agree well between data and MC simulation, as shown in Table~\ref{tab:yield}.
The combination method is cross-checked by performing the unfolding in each channel individually to
the $\ell$+jets phase space described in Sec.~\ref{sec:fiducialregion} and comparing these alternative cross-section estimates with
the one based on the combined data. The final results are found to be
consistent. 

\subsection{Particle- and parton-levels fiducial region definitions}
\label{sec:fiducialregion}

Particle-level corrections to the data are derived from leptons and jets in
simulated \ttbar\ events that are constructed using stable particles, with a mean
lifetime greater than $0.3 \times
10^{-10}$ seconds, 
which result directly from the hard-scattering 
$pp$ interaction or from subsequent decays of particles with a shorter
lifetime.  

All leptons
($e$, $\mu$, $\nu_{e}$ ,$\nu_\mu$, $\nu_{\tau}$) not from hadron
decays are considered as prompt isolated leptons.
The leptons from $\tau$ decays are accepted only if the parent $\tau$
is not a hadron decay product itself.  
The four-momenta of photons within a cone of $\Delta R$ = 0.1 around the
electron or muon direction are added to those of the leptons (dressed leptons). 
Both the small-$R$ and large-$R$ jets are reconstructed using all stable particles
except for the selected dressed leptons.
The trimming procedure applied to detector-level jets is also
applied to particle-level jets. 
A small-$R$ jet with \pt\ $>$ 25 \GeV\ and $|\eta| < 2.5$ is considered
to be ``$b$-tagged''  if there exists at least one $b$-hadron with $\pt > 5$
\GeV\ clustered in the
jet.\footnote{The $b$-hadrons are not stable and do not contribute to
  the total four-vector of the jet, only their decay products
  do. However, they are clustered with their energy set to a negligible
  value to check that they match the jet geometrically~\cite{Cacciari:2008gn}.}

The missing transverse momentum $\met$  is the magnitude of the vector
sum of the momenta of neutrinos not resulting from hadron decays.

To minimize the theoretical input to the measurement, the fiducial
region is chosen to follow the detector-level event selections closely,
including the kinematic requirements on the objects and the requirements
on the event topology.
In contrast to the detector-level selection, no overlap removal
procedure is applied to the leptons and jets, 
and no isolation requirement is imposed on the leptons.
Using the particle-level objects defined above, the fiducial region is defined by requiring:
\begin{itemize}
  \item Exactly one lepton (electron or muon) with $\pt$
    $>$ 25 \GeV, $|\eta|<$2.5.
  \item \Etmiss $>$ 20 \GeV\ and \Etmiss+\mtw $>$ 60 \GeV.
  \item At least one small-$R$ jet with $\pt>$ 25 \GeV, 
    $|\eta|$ $<$ 2.5, and a distance $\Delta R<$1.5 from the
    lepton. If there is more than one such jet, the one with the
    largest \pt\ is considered to be the leptonic $b$-jet candidate (the $b$-jet associated to the leptonic top quark decay).
  \item At least one trimmed large-$R$ jet with $\pt$$>$ 300
    \GeV, mass $>$ 100 \GeV, $\sqrt{d_{12}}>$ 40 \GeV, and $|\eta|$ $<$
    2, well separated from both the lepton ($\Delta\phi$ $>$ 2.3) and the
    leptonic $b$-jet candidate ($\Delta R >$ 1.5).
    The jet mass is reconstructed from the four-vector sum of the particles
    constituting the jet. If more than one large-$R$ jet satisfies these criteria, 
    the one with largest \pt\ is chosen. The jet passing this
    selection is referred to as the particle-level top-jet candidate.
\item  At least one $b$-tagged small-$R$   jet such that $\Delta R(
  \mathrm{jet}_{R=1.0}, \mathrm{jet}_{R=0.4}) < 1$ and/or the leptonic
  $b$-jet candidate is $b$-tagged.
\end{itemize}
The particle-level event selection is summarized in Table 1.
Fiducial particle-level corrections are determined by using only simulated \ttbar\ events in which
exactly one of the $W$ bosons, resulting from the decay of the \ttbar\
pair, decays to an electron or a muon either directly or through a
$\tau$ lepton decay. All other \ttbar\ events are not used. 
The cross-section is then determined as a function of the 
particle-level top-jet candidate transverse
momentum,~\ptptcl.

For the parton level, the top quark that decays
to a hadronically decaying $W$ boson is considered just before the decay and
after QCD radiation,
selecting events in which the momentum of such a top quark, 
\ptptn, is larger than 300 \GeV.
Parton-level corrections are determined by using only simulated \ttbar\ events in which
exactly one of the $W$ boson decays to an electron or a muon or a $\tau$
lepton (including hadronic $\tau$ decays). The correction to the full
parton-level phase space defined above is obtained by accounting for
the branching ratio of \ttbar\ pairs to the \ljets channel.

\subsection{Unfolding to particle  and parton levels}
\label{sec:subunfolding}

The procedure to unfold the distribution of \ptreco, the \pt\ of the detector-level
leading-$\pt$ trimmed large-$R$ jet, 
 to obtain the differential cross-section as a function of \ptptcl\ is
 composed of several steps, outlined in: 
\begin{eqnarray}
 \frac{\dsigma_{\ttbar}}{d\ptptcl} (\ptptcl^{i}) &=&\frac{N^{i}_{\mathrm{ptcl}}}{\Delta  \ptptcl^{i} \lum} 
\nonumber \\
&=& \frac{1}{\Delta
  \ptptcl^{i}\lum f_{\mathrm{ptcl!reco}}^i } \cdot \sum_{j} M_{ij}^{-1} f_{\mathrm{reco!ptcl}}^{j}
 f_{\ttbar \mathrm{, \ell+jets}}  (N^{j}_{\mathrm
  {reco}}-N^{j}_{\mathrm{reco,bgnd}}),
\label{formula:unfold_part} 
\end{eqnarray}
where $N^{j}_{\mathrm{reco}}$
is the number of observed events in bin $j$ of \ptreco\ with the
detector-level selection applied,
$N^{i}_{\mathrm{ptcl}}$ is the total number of events in bin $i$ of
\ptptcl\ that meet the fiducial region selection, 
$\Delta  \ptptcl^{i}$ is the size of bin $i$ of 
  \ptptcl, and $\lum$  is the integrated luminosity of the 
  data sample. The corrections that are applied to \ptreco\ are
 all extracted from
the nominal {\sc Powheg+Pythia} $\ttbar$ sample. 

First, the non-$\ttbar$ background contamination,
$N^j_{\mathrm{reco,bgnd}}$, 
is subtracted from the observed number of events
in each \ptreco\ bin.
  The contribution from non-{\lplus}  \ttbar\ events is taken into account by the multiplicative
  correction $f_{\ttbar \mathrm{, \ell+jets}}$, which represents the 
  fraction of $\ell$+jets $\ttbar$ events extracted from the nominal {\sc Powheg+Pythia} $\ttbar$ sample.

\begin{figure*}
\centering
\subfigure[\hspace*{0.15cm} ]{\includegraphics[width=0.48\textwidth]{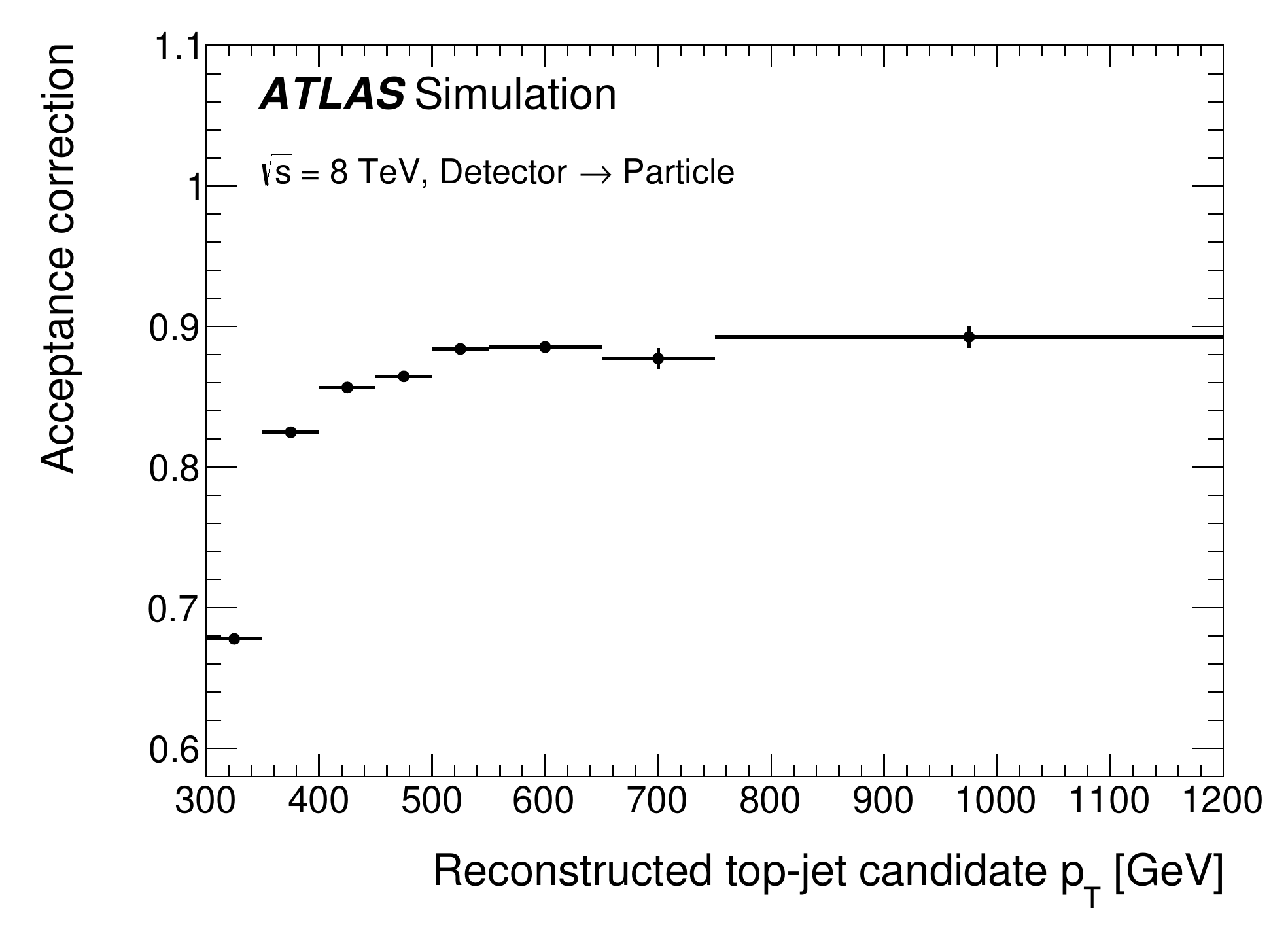}}
\subfigure[\hspace*{0.15cm} ]{\includegraphics[width=0.48\textwidth]{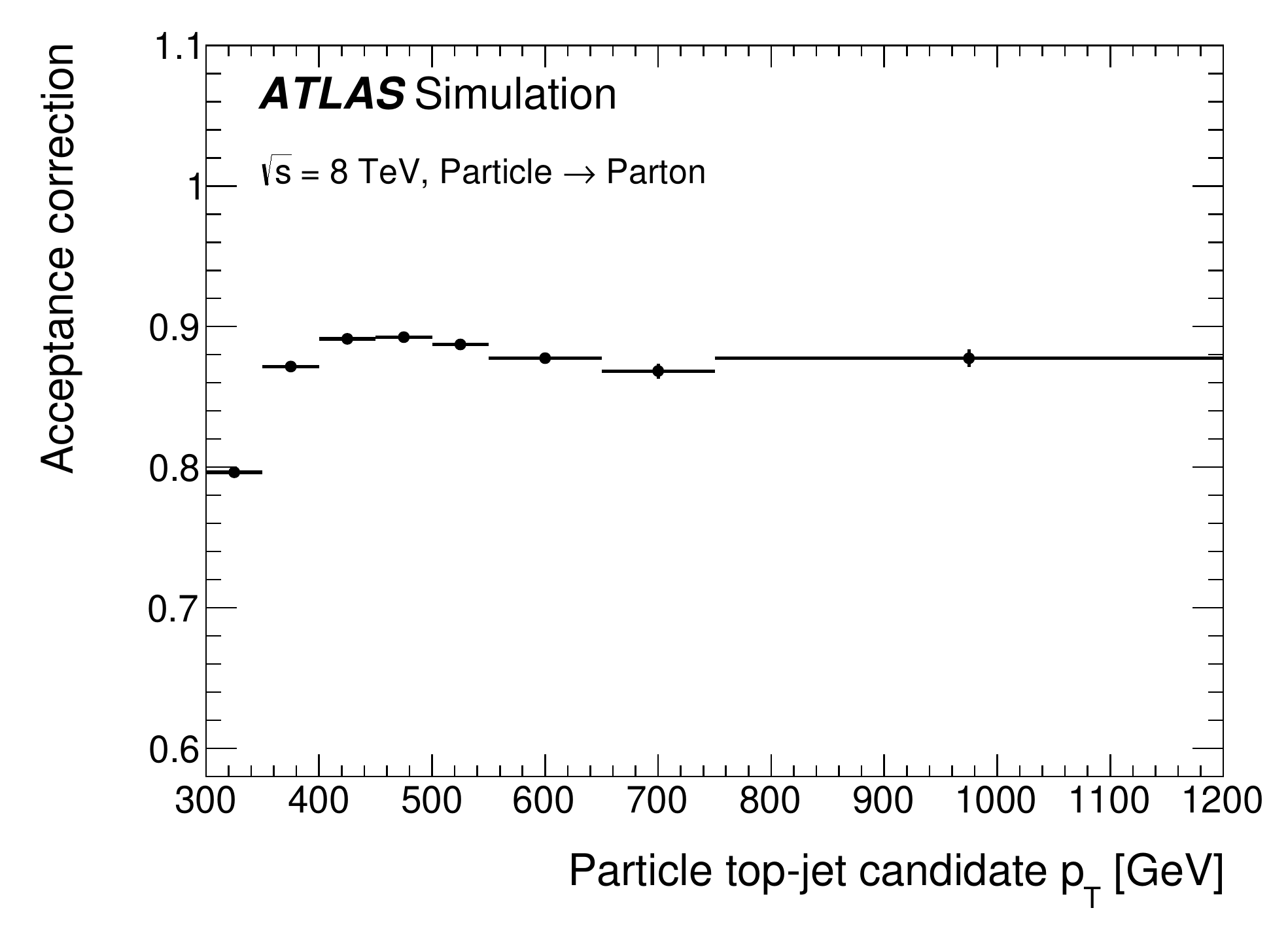}}

  \caption{ (a) Distribution of the correction factor
    $f_{\mathrm{reco!ptcl}}$  as a function of \ptreco.
    It represents the ratio of the number of events that meet both the
    detector-level and particle-level
    to the number of events that satisfy the detector-level selection requirements. 
   (b) Distribution of the correction factor
   $f_{\mathrm{ptcl!parton}}$  as a function of \ptptcl.
   It represents the ratio of the number of events that meet both the
   parton-level and particle-level 
   to the number of events that satisfy only the particle-level selection requirements. 
}
  \label{fig:fakecorr}
\end{figure*}

In a second step the correction factor $f_{\mathrm{reco!ptcl}}^{j}$,
also referred to as acceptance correction, corrects
the \ptreco\ spectrum for the $\ttbar$ events
that pass the detector-level selection but fail the
particle-level selection. 
For each \ptreco\ bin $j$, $f_{\mathrm{reco!ptcl}}^{j}$ is defined as the ratio
of the number of events that meet both the detector-level and particle-level
selections to
the number of events that satisfy the detector-level selection.
The distribution of the acceptance correction $f_{\mathrm{reco!ptcl}}^j$ is shown in
Fig.~\ref{fig:fakecorr}(a) for various MC generators.

\begin{figure*}
\centering
\subfigure[\hspace*{0.15cm} ]{\includegraphics[width=0.48\textwidth]{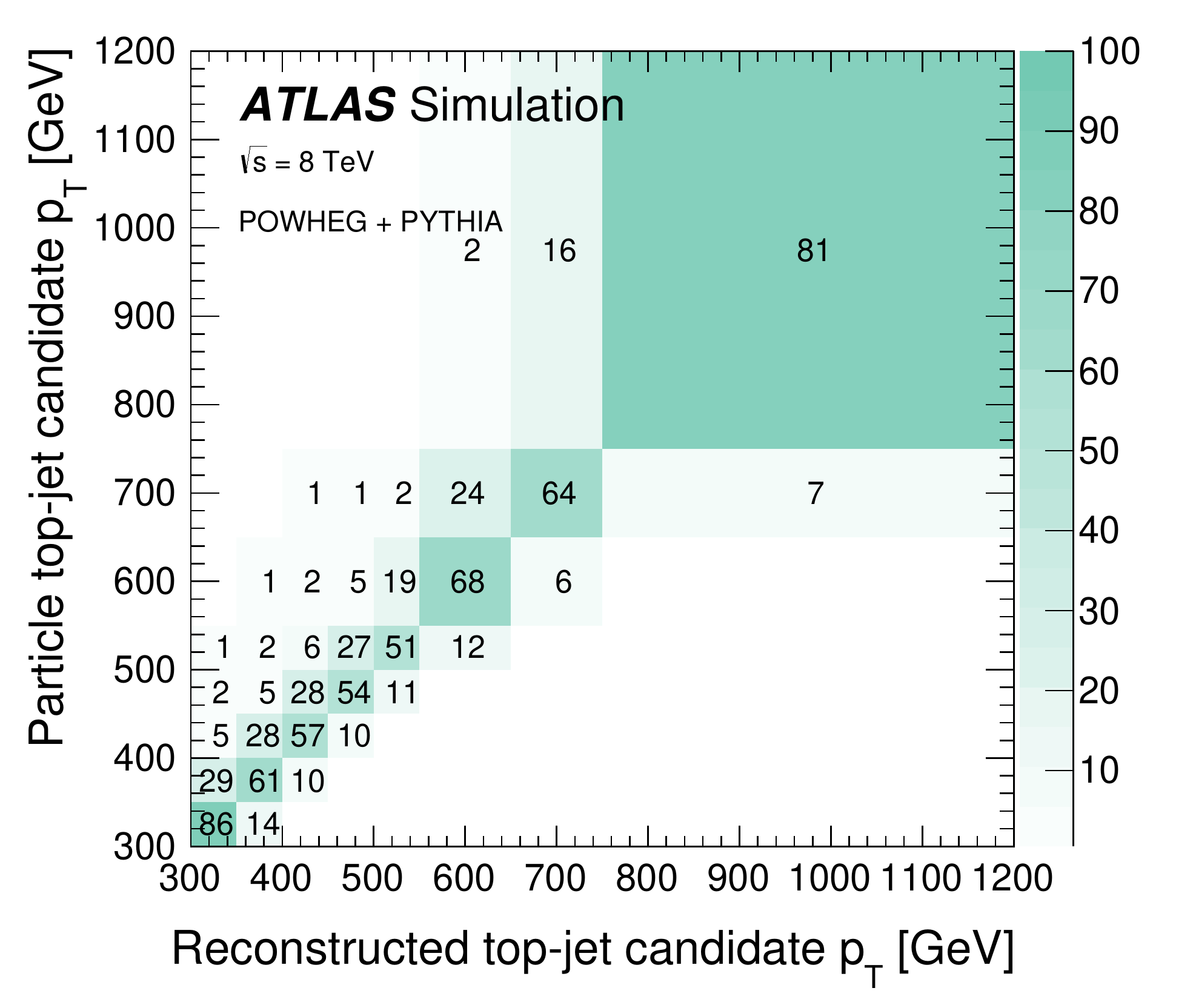}}
 \subfigure[\hspace*{0.15cm} ]{\includegraphics[width=0.48\textwidth]{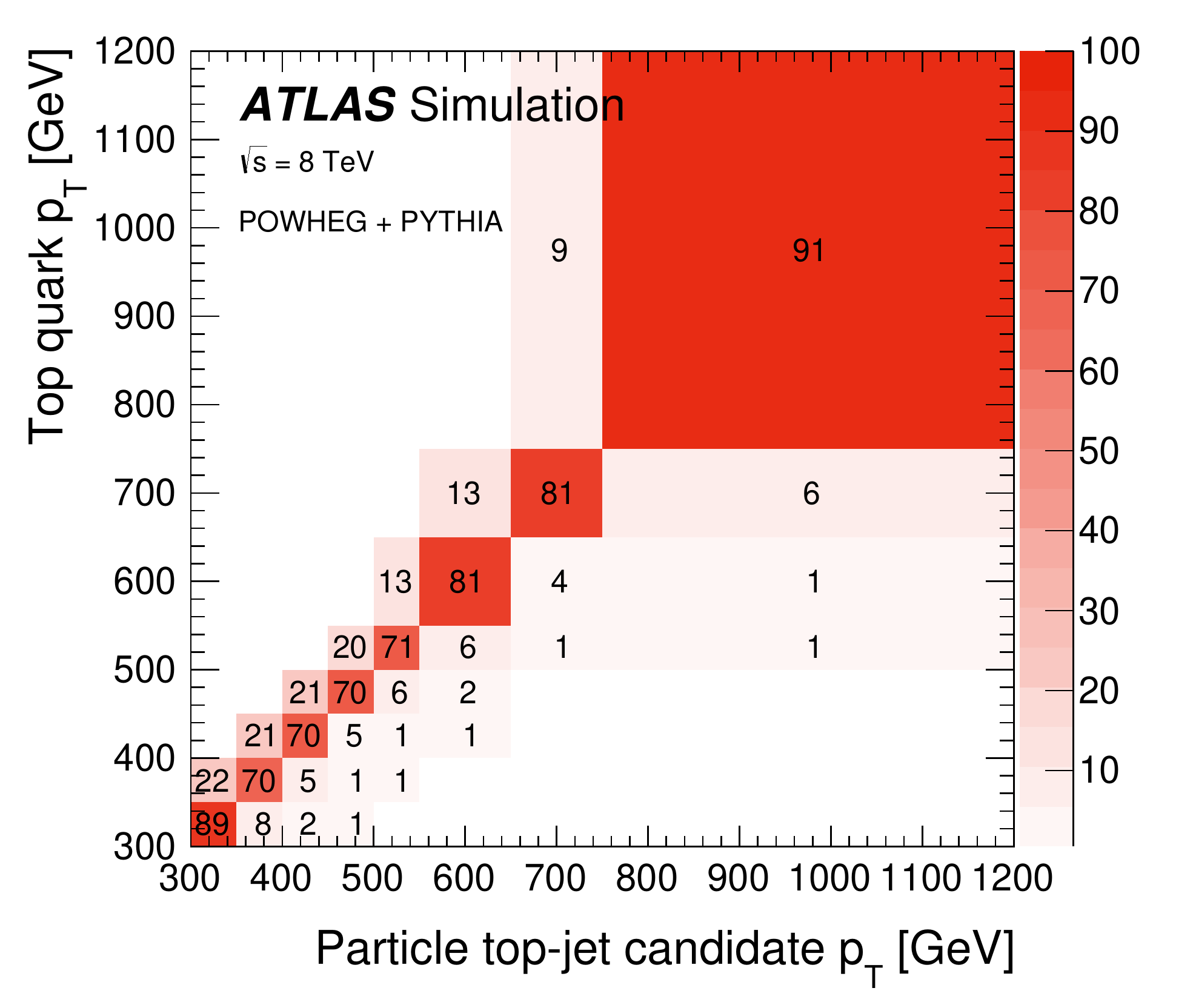}}
  \caption{ (a) Migration matrix between the particle-level \ptptcl and
    reconstructed detector-level \ptreco. (b) Migration matrix between the generated \ptptn and the particle-level \ptptcl. 
   The unit of the matrix elements is the probability (expressed in percentage) for an event generated
    at a given value to be reconstructed at another value (each row
    adds up to 100\%). }
\label{fig:migration}
\end{figure*}

The third step corrects for detector resolution effects.
A migration matrix is constructed to correlate the \ptreco-binned 
distribution to the \ptptcl distribution. 
The matrix $M_{ij}$ represents the probability
for an event with  \ptptcl\ in bin $i$  to have a \ptreco\ in bin $j$.
This matrix is shown in Fig.~\ref{fig:migration}(a). 
It shows that approximately 50\% to
 85\% of events have values of \ptptcl\ and of \ptreco\ that fall in
 the same bin.

The inversion of the migration matrix to correct \ptreco\ to the
particle level
 is carried out by an unfolding scheme based on Tikhonov
 regularization which is implemented through the singular value
 decomposition (SVD) of the matrix~\cite{svd}.  
This scheme is chosen to reduce sizable statistical fluctuations that are
introduced by instabilities in the inversion procedure. 
The unfolding regularization parameter, which characterizes
the size of the expansion of the solution to the inversion problem,
is optimized according to the procedure described in
Ref.~\cite{svd}. 
In parallel the bin size for the \ptptcl\ (and \ptreco) 
distribution is optimized such that systematic uncertainties are larger than statistical uncertainties in each
bin, and such that the width of each bin corresponds to at least one and a half times the expected resolution in that bin.
 The former
requirement is introduced to minimize statistical fluctuations when
estimating systematic uncertainties.  The typical expected fractional
resolution for \ptreco\ in \ttbar\ simulated events ranges from
7\%   to  3\% for \ptreco\ values between 250 \GeV\ and 1.2 \TeV.
Finally, the optimization requires the unfolding to be unbiased,
i.e., that a given input \ptptcl\ spectrum is recovered on average
by the unfolding procedure. 
After rounding to the nearest 50 \GeV, this procedure
results in bin widths of 50 \GeV\ between 300 \GeV\ and 550 \GeV, 100 \GeV\ between
550 \GeV\ and 750 \GeV, while the last bin spans 750 \GeV\ to 1200 \GeV.
 Just one event with reconstructed
$\pt$ = 1535 \GeV\ falls outside this region in the \muplus\ sample, and none
in the \eplus\ sample.

\begin{figure*}
\centering
\subfigure[\hspace*{0.15cm} ]{\includegraphics[width=0.48\textwidth]{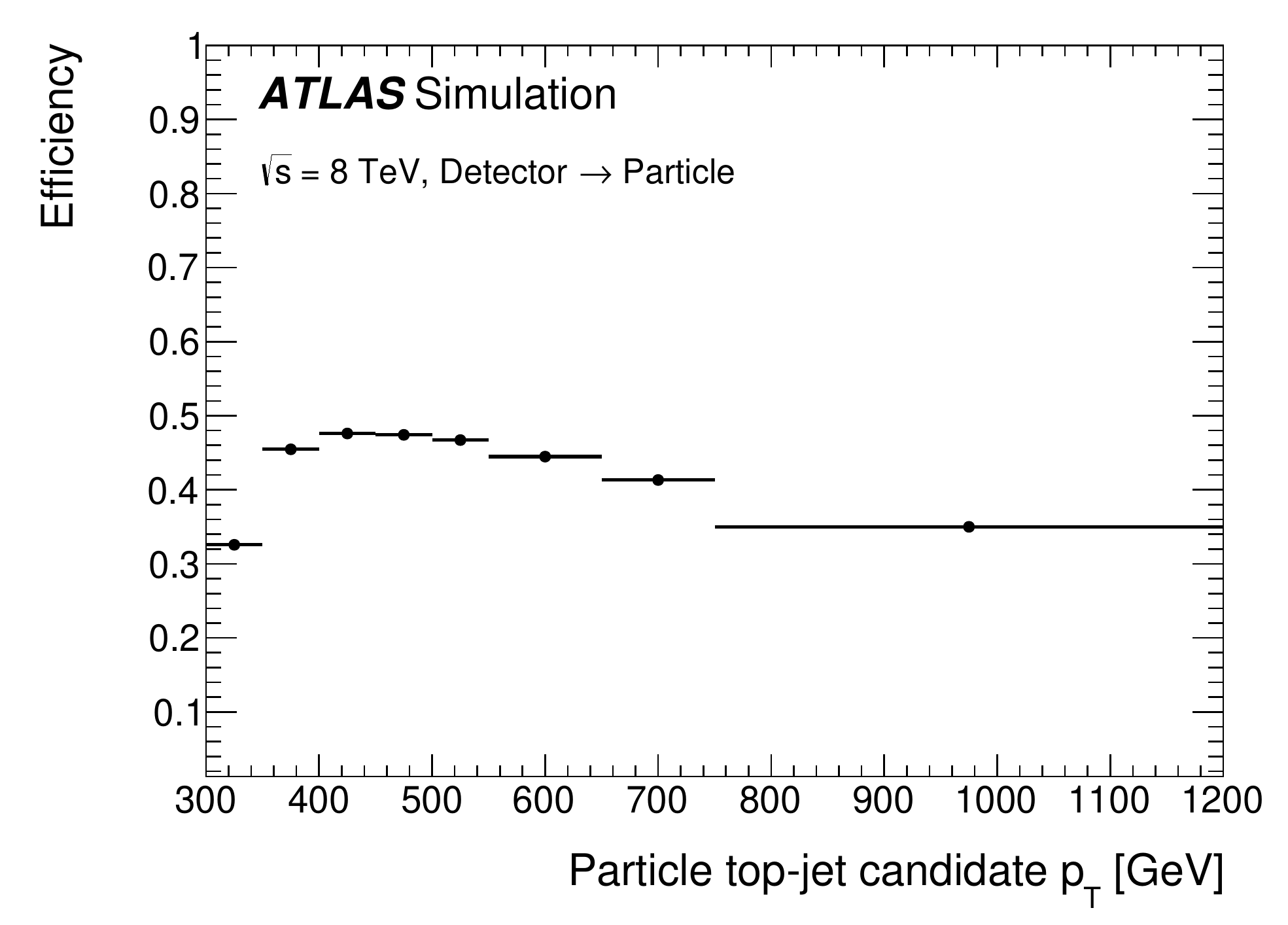}}
\subfigure[\hspace*{0.15cm} ]{\includegraphics[width=0.48\textwidth]{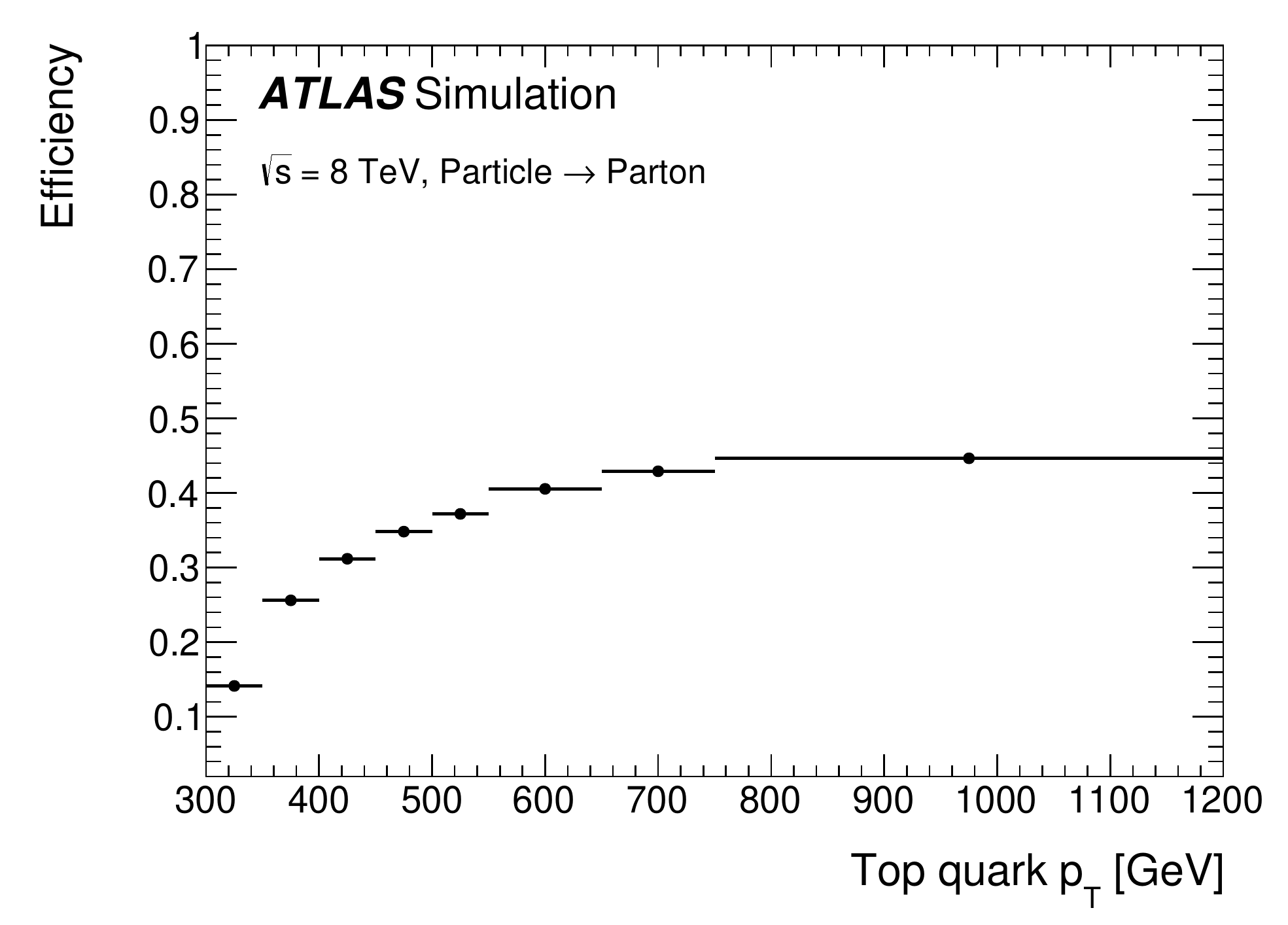}}
  \caption{(a) Distribution of the correction factor
    $f_{\mathrm{ptcl!reco}}$  as a function of \ptptcl.
    It represents the ratio of  events that meet both the particle-level and detector-level 
    to those that satisfy the  particle-level selection requirements. 
    (b) Distribution of the correction factor
    $f_{\mathrm{parton!ptcl}}$ as a function of \ptptn.  
    It represents the ratio of  events that meet both the parton-level and particle-level 
    to those that satisfy the  parton-level selection requirements. 
 }
  \label{fig:eff}
\end{figure*}

The fourth step is to apply a bin-by-bin correction factor $f_{\mathrm{ptcl!reco}}^{i}$,
also referred to as efficiency correction,  which restores the
contribution of  \ttbar\ events that fulfill the particle-level
selection but not the detector-level selection. 
This factor is defined as the ratio of the number of events that satisfy both the particle-level and
detector-level selections to the number that meet the
selection at particle level only.
The distribution of the efficiency correction
$f_{\mathrm{ptcl!reco}}^{i}$  is shown in Fig.~\ref{fig:eff}(a).

The  ability of the full correction procedure to recover a distribution that is
significantly different from the nominal $\ttbar$ sample is tested.
Simulated \ttbar\ events are reweighted such that the \ptreco\
distribution 
matches the data.
The corresponding \ptptcl\ spectrum of the distorted \ptreco\ input spectrum is
recovered with subpercent accuracy after unfolding.

The differential cross-section as a function of \ptptn\ is then derived according to:

\begin{eqnarray}
\frac{\dsigma_{\ttbar}}{d\ptptn} (\ptptn^{k})
 &=&\frac{N^{k}_{\mathrm{parton}}}{\mathcal{B} \Delta  \ptptn^{k}  \lum} 
\nonumber \\
&=& \frac{1}{\mathcal{B} \Delta
  \ptptn^{k}  \lum  f_{\mathrm{parton!ptcl}}^k } \cdot \sum_{j} \hat{M}_{jk}^{-1}  f_{\mathrm{ptcl!parton}}^{j}
 N^{j}_{\mathrm
  {ptcl}}.
\label{formula:unfold_gen} 
\end{eqnarray} 
Similarly to Eq.~(\ref{formula:unfold_part}), $N^{j}_{\mathrm{ptcl}}$ is the total number of events in bin $j$ of
\ptptcl\
 that enter the particle-level fiducial region described in Sec.~\ref{sec:fiducialregion}, $N^{k}_{\mathrm{parton}}$
is the number of events in bin $k$ of \ptptn\ in the full phase space, $\Delta
  \ptptn^{k}$ is the size of bin $k$ of the parton-level
  \ptptn\ (and of \ptptcl), $\lum$  is the total integrated
  luminosity of the data sample, and $\cal{B}$=0.438~\cite{BRRef} is the branching ratio for
  \ttbar\ events with exactly one of the $W$ bosons, from the decay of the \ttbar\
pair, decaying to an electron or a muon or a $\tau$ lepton.

 The corrections that are applied to the  \ptptcl\ variable are derived
 following steps similar to the ones described to derive
$\dsigma_{\ttbar}/d\ptptcl$. They are also extracted from the nominal
 {\sc Powheg+Pythia} $\ttbar$ sample.
First, the factor $f_{\mathrm{ptcl!parton}}^{j}$ corrects
the \ptptcl\ spectrum for the $\ttbar$ events
that pass the particle-level selection but
fail the parton-level selection, 
 shown in
Fig.~\ref{fig:fakecorr}(b).
Effects relating \ptptn\ to \ptptcl  are corrected with the same matrix unfolding procedure used for detector effects. 
This migration matrix $\hat{M}_{jk}$ is shown in
Fig.~\ref{fig:migration}(b). 
A final correction factor $f_{\mathrm{parton!ptcl}}^{k}$ is
applied in bins of \ptptn\ to correct the result from the
particle level to the partonic phase space, 
 shown in Fig.~\ref{fig:eff}(b).

To test the two-step derivation, the cross-section
is also obtained by directly correcting the reconstructed distribution
to parton level in a single step. The results are found to be
consistent.

\subsection{Propagation of statistical and systematic uncertainties}
\label{sec:subsyst}

The propagation of  statistical and systematic uncertainties is performed in
the same way for both the particle-level and parton-level results.
The impact of the data statistical uncertainty is evaluated by performing 1000
pseudoexperiments in which independent Poisson fluctuations in each \ptreco\ bin are
assumed. The statistical uncertainty due to the limited size of the signal and background MC samples used to
correct the data are estimated by performing 1000 pseudoexperiments using
the bootstrap method~\cite{bootstrap}, 
which builds 1000 statistically connected
(co-varied) replicas of individual simulated signal or background spectra and
derives the associated corrections.

For each systematic uncertainty arising from detector modeling,
background modeling, and the electroweak
correction factor,
a varied \ptreco\ distribution is obtained and unfolded using
corrections extracted from the nominal signal and background
samples. 
The correlation between each systematic uncertainty's effect on the signal and background spectra is taken into account.
For the  \ttbar\ generator, parton shower, and ISR/FSR uncertainties, a systematic uncertainty variation is defined as the difference between
the generated and unfolded cross-section of a given generator, with
unfolding corrections extracted with an alternative generator (or alternative
generator setting). 
The PDF uncertainty is computed by unfolding the nominal sample with
correction factors extracted by 
reweighting the nominal sample at the hard-process level for each
variation of the PDF.

\begin{figure*}
 \centering
  \subfigure[\hspace*{0.15cm} ]{\includegraphics[width=0.49\textwidth]{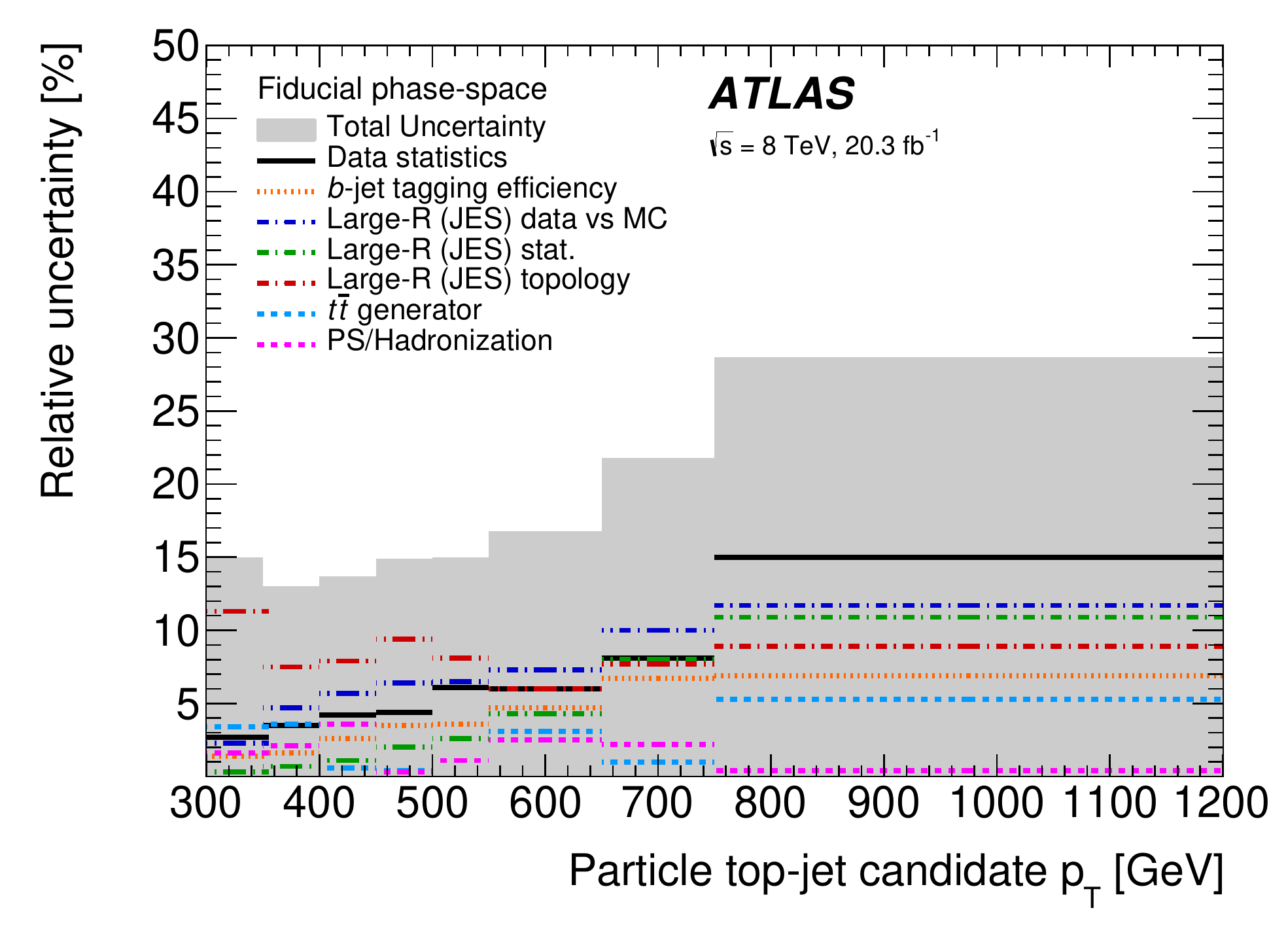}}
  \subfigure[\hspace*{0.15cm} ]{\includegraphics[width=0.49\textwidth]{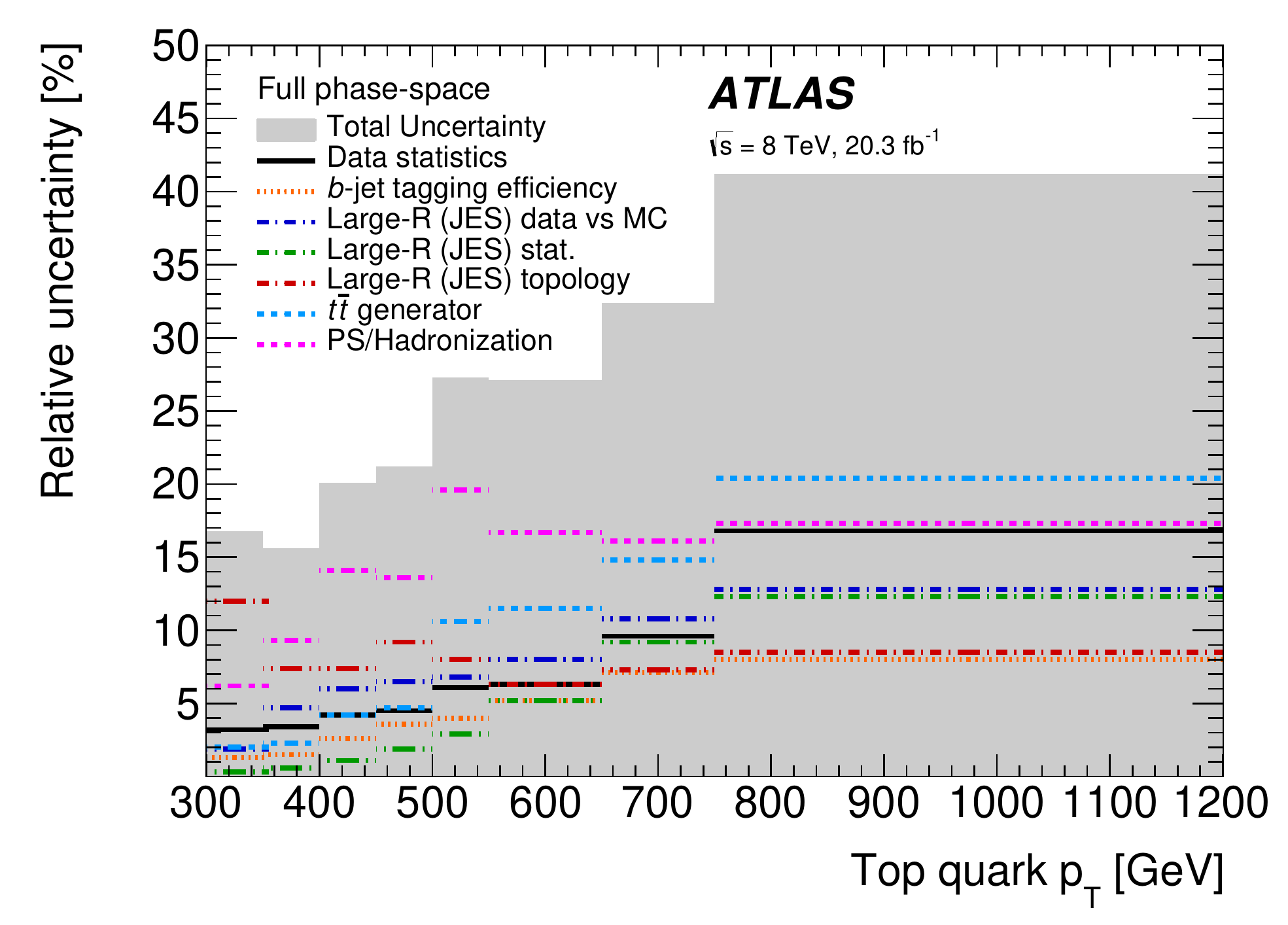}}
  \caption{ Relative uncertainties on
    (a) the particle-level differential cross section $\dsigma_{t\bar{t}}/d\ptptcl^{i} $
    and (b) the parton-level differential cross section $\dsigma_{t\bar{t}}/d\ptptn^{i} $. The total
    uncertainty (band) is shown along with the effect of the dominant
    uncertainties. The components ``Large-$R$ (JES) stat.'' and
    ``Large-$R$ (JES) data vs MC'' are, respectively, the statistical
    uncertainty and the systematic uncertainty associated with the
    difference in jet response between data and MC simulation when balancing $\pt$ in photon+jet events.
}
  \label{fig:syst}
\end{figure*}

Figure ~\ref{fig:syst} shows  the effect of the statistical and systematic uncertainties on $\dsigma_{\ttbar}/d\ptptcl$ and
$\dsigma_{\ttbar}/d\ptptn $. 
The total uncertainty generally increases
with the measured $\pt$  and ranges from 
13\% to 29\% for the particle-level cross-section, and from
15\% to 41\% for the parton-level cross-section.
The dominant uncertainty for the
particle-level cross-section is the large-$R$
jet energy scale, in particular its components due to 
the topology uncertainty at low $\pt$ and the uncertainty from $\pt$ balance in photon+jet
events at high $\pt$. The experimental uncertainties have
a comparable size at parton level. However, the reported parton-level
cross-section has significantly larger systematic 
 uncertainties than the particle-level cross-section since it is
 affected by larger $\ttbar$ modeling
 uncertainties.
 The parton shower or generator uncertainties are dominant for nearly all
 $\pt$ bins of the parton-level cross-section, which illustrates the
 benefit of defining a particle-level cross-section in a fiducial
 region closely following the detector-level selection. A detailed
 breakdown of the systematic uncertainties is provided in Appendix
 \ref{Sec:Appendix:Syst}.

A covariance matrix including the effect of all uncertainties is
calculated at particle level to make quantitative comparisons with
theoretical predictions.
This covariance matrix is obtained by summing two covariance matrices.
The first covariance matrix incorporates uncertainties from detector and background
modeling by performing 250,000
pseudoexperiments.
 In each pseudoexperiment, the data~\ptreco\ distribution is varied
 following a Poisson distribution.
Gaussian-distributed shifts are coherently added for each systematic
uncertainty effect  by scaling each Poisson-fluctuated bin with the 
 relative variation from the associated systematic uncertainty effect.
Differential cross-sections are obtained by unfolding each varied~\ptreco\ distribution with the nominal
corrections, and the results are used to compute a covariance matrix.
 
The second covariance matrix
 is obtained by summing  four separate covariance matrices corresponding
 to the  effects of \ttbar\ generator, parton shower,
 ISR/FSR, and PDF uncertainties.  The standard deviations of the covariance matrices
are derived by scaling the
measured cross-section with the appropriate relative systematic
uncertainty. The bin-to-bin correlation value is set to unity for the
generator, parton shower, and ISR/FSR matrices, while it
is set to 0.5 for the PDF matrix. This value is
motivated by the fraction of the bins in which a single PDF set dominates in the
determination of the envelopes used for their respective estimates.
The procedure for these signal modeling uncertainties is needed because these effects cannot be represented by a variation at
the detector level, and so cannot be included in the
pseudoexperiment formalism used to build the first covariance
matrix.

The correlation matrix derived from the particle-level
covariance matrix is shown
in~Table~\ref{tab:CorrPtptcl}.
Agreement between the measured differential cross-sections and
various predictions
is quantified by calculating $\chi^2$ values employing
the covariance matrix and by inferring
corresponding $p$-values.
The  $\chi^2$ are evaluated using: 
\begin{linenomath}
\begin{equation*}
\chi^2 = V^{\rm T} \cdot {\rm Cov}^{-1} \cdot V,
\end{equation*} where $V$ is the vector of differences between measured
differential cross-section values and predictions, and  ${\rm
  Cov}^{-1}$  is the inverse of the covariance matrix.
\end{linenomath}

\input{corr_ptptcl}

%% file: corr_ptptcl.tex
\begin{table*}[htbp]
\begin{center}
\noindent\makebox[\textwidth]{
\begin{tabular}{r|rrrrrrrr}
\hline
\ptptcl [\GeV] & 300--350 & 350--400 & 400--450 & 450--500 & 500--550 & 550--650 & 650-750 & 750-1200   \\  \hline
300--350  & 1.00 & 0.83 & 0.79 & 0.79 & 0.72 & 0.63 & 0.58 & 0.51 \\
350--400  & 0.83 & 1.00 & 0.83 & 0.80 & 0.76 & 0.74 & 0.67 & 0.60 \\
400--450  & 0.79 & 0.83 & 1.00 & 0.87 & 0.79 & 0.78 & 0.77 & 0.63 \\
450--500  & 0.79 & 0.80 & 0.87 & 1.00 & 0.89 & 0.76 & 0.77 & 0.66 \\
500--550  & 0.72 & 0.76 & 0.79 & 0.89 & 1.00 & 0.84 & 0.75 & 0.62 \\
550--650  & 0.63 & 0.74 & 0.78 & 0.76 & 0.84 & 1.00 & 0.89 & 0.71 \\
650--750  & 0.58 & 0.67 & 0.77 & 0.77 & 0.75 & 0.89 & 1.00 & 0.87 \\
750--1200  & 0.51 & 0.60 & 0.63 & 0.66 & 0.62 & 0.71 & 0.87 & 1.00 \\
\hline
\end{tabular}
}
\caption{Correlation matrix between the bins of the particle-level differential cross-section as a function of \ptptcl.}
\label{tab:CorrPtptcl}
\end{center}
\end{table*}

%% file: results.tex
\section{Results and interpretation}
\label{sec:results}

\begin{figure*}
\centering
  \includegraphics[width=0.7\textwidth]{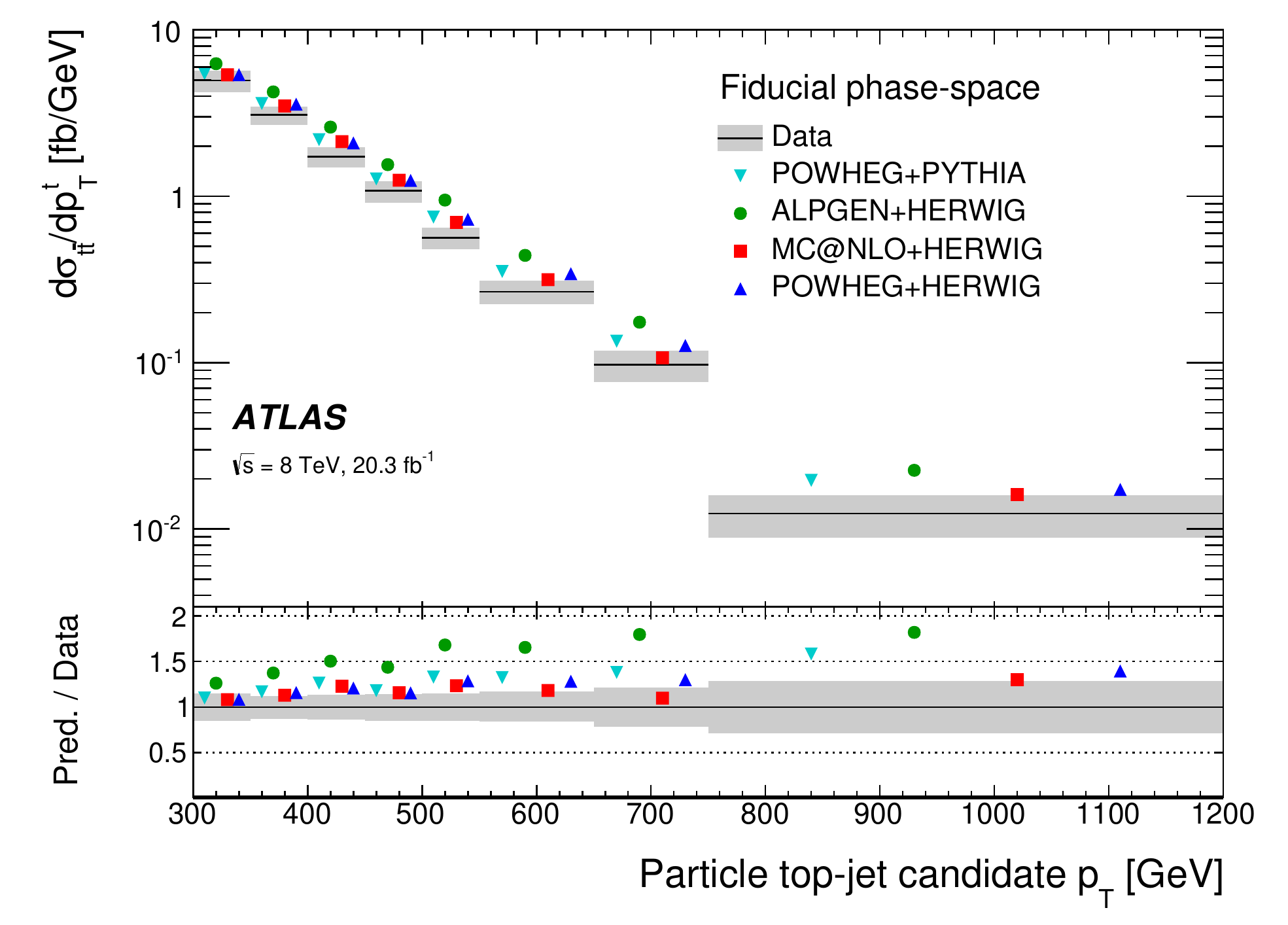}
  \caption{Fiducial particle-level differential cross-section as a
    function of the hadronic top-jet candidate $\pt$. 
    {\sc Powheg+Pythia}, {\sc Powheg+Herwig}, {\sc MC@NLO+Herwig}, and {\sc Alpgen+Herwig} predictions are 
    compared with the final results. 
          MC samples are normalized to the NNLO+NNLL inclusive cross-section $\sigma_{t\bar{t}}= 253$~pb.
	  No electroweak corrections are applied to the predictions.
          The lower part of the figure shows the ratio of
          the MC prediction to the data. 
          The shaded area includes the total statistical plus
          systematic uncertainties. The points of the various predictions are spaced along
        the horizontal axis for presentation only; they
        correspond to the same $\pt$ range.}
  \label{fig:results:particle}
\end{figure*}

\begin{figure*}
\centering
  \includegraphics[width=0.7\textwidth]{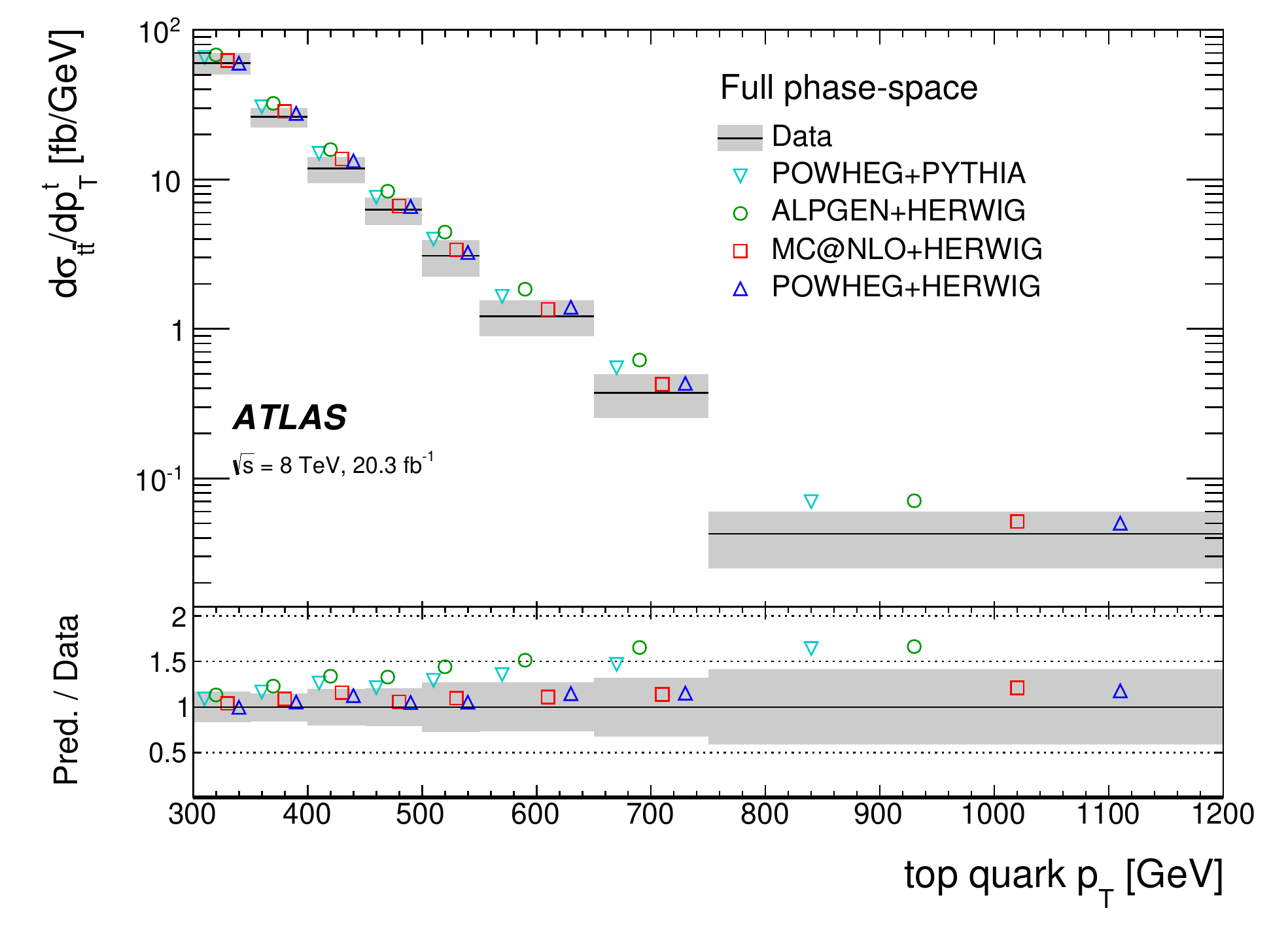}
  \caption{Parton-level differential cross-section as a
    function of the hadronically decaying top quark $\pt$. 
    {\sc Powheg+Pythia}, {\sc Powheg+Herwig}, {\sc MC@NLO+Herwig}, and {\sc Alpgen+Herwig} predictions are 
    compared with the final results.  
          MC samples are normalized to the NNLO+NNLL inclusive cross-section $\sigma_{t\bar{t}}= 253$~pb.
	  No electroweak corrections are applied to the predictions.
          The lower part of the figure shows the ratio of
          the MC prediction to the data. 
	  The shaded area includes the total statistical plus systematic uncertainties.  The points of the various predictions are spaced along
        the horizontal axis for presentation only; they
        correspond to the same $\pt$ range.}
  \label{fig:results:parton}
\end{figure*}

The unfolding procedure is applied to the
observed top-jet candidate $\pt$ distribution. The cross-sections are provided in Table~\ref{tab:combo2}
and Fig.~\ref{fig:results:particle} for the particle-level cross-section, and in Table~\ref{tab:comboparton2}
and Fig.~\ref{fig:results:parton} for the parton-level
cross-section. 
The higher efficiency of reconstruction techniques for highly boosted top quarks allows measurement of 
the top quark
$\pt$ spectrum up to 1200 \GeV.
The differential cross-section is measured 
over two orders of magnitude. 
The measured differential cross-sections are compared to
 the predictions from {\sc Alpgen}+{\sc Herwig}, {\sc MC@NLO}+{\sc Herwig}, {\sc Powheg}+{\sc Herwig},  and
{\sc Powheg}+{\sc Pythia} $\ttbar$ samples normalized to the NNLO+NNLL inclusive
cross-section. The electroweak corrections are not applied to the
{\sc Powheg}+{\sc Pythia} prediction in these figures in order to compare it on an equal footing
with the other generators.
All generators produce a top quark $\pt$
spectrum that is harder than the one observed,  
with a difference that generally increases with $\pt$. 
The MC prediction to data ratio is approximately the same at both the 
particle and parton levels for {\sc Powheg}+{\sc Pythia}, which was
used to extract the unfolding corrections. 
However, it changes significantly when going from particle level to
parton level for the other MC generators, in particular for
{\sc Powheg}+{\sc Herwig}, and {\sc Alpgen}+{\sc Herwig}, due to the different
parton-level corrections in these MC generators. 
The level of agreement is better at parton level than at particle level because 
the parton level is affected by larger systematic uncertainties.

\begin{table}
\centering
\input{sum_pt_hadTop_SVD_particle__ResultTable.tex}

\caption{Fiducial particle-level differential cross-section, with statistical and systematic uncertainties, as a
    function of the top-jet candidate~$\pt$.}
\label{tab:combo2}
\end{table}

\begin{table}
\centering
\input{sum_pt_hadTop_SVD_parton__ResultTable.tex}

\caption{Parton-level differential cross-section, with statistical and systematic uncertainties, as a
    function of the hadronically decaying top quark~$\pt$.}
\label{tab:comboparton2}
\end{table}

The $\chi^2$ and $p$-values that quantify the level of agreement
between the particle-level predictions and data are listed in Table~\ref{tab:pvalues_combined}. 
Within uncertainties, the 
differences are not significant for {\sc Powheg}+{\sc Pythia}, {\sc
  Powheg}+{\sc Herwig} and {\sc MC@NLO}+{\sc Herwig}, for which
$p$-values of 0.11 (for {\sc Powheg}+{\sc Pythia} without
electroweak corrections), 0.41, and 0.14 are obtained, respectively. Only the
prediction of
{\sc Alpgen}+{\sc Herwig} is significantly disfavored by the data at the
particle level with a $p$-value
of $5.9\cdot10^{-5}$. 

\input{pvaluesTableCombined}

\begin{figure*}
\centering 
  \subfigure[\hspace*{0.15cm} ]{\includegraphics[width=0.49\textwidth]{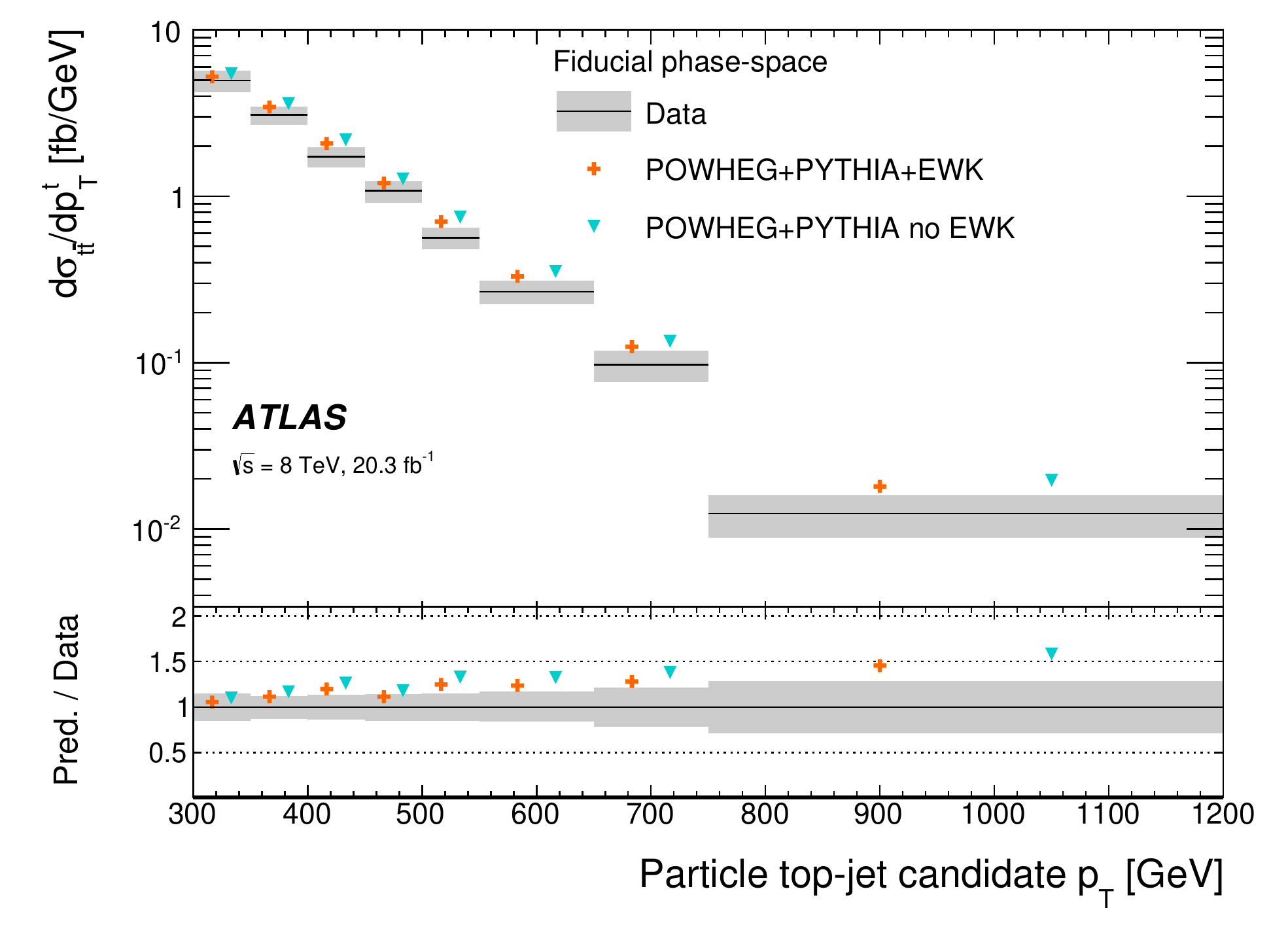}}
  \subfigure[\hspace*{0.15cm} ]{\includegraphics[width=0.49\textwidth]{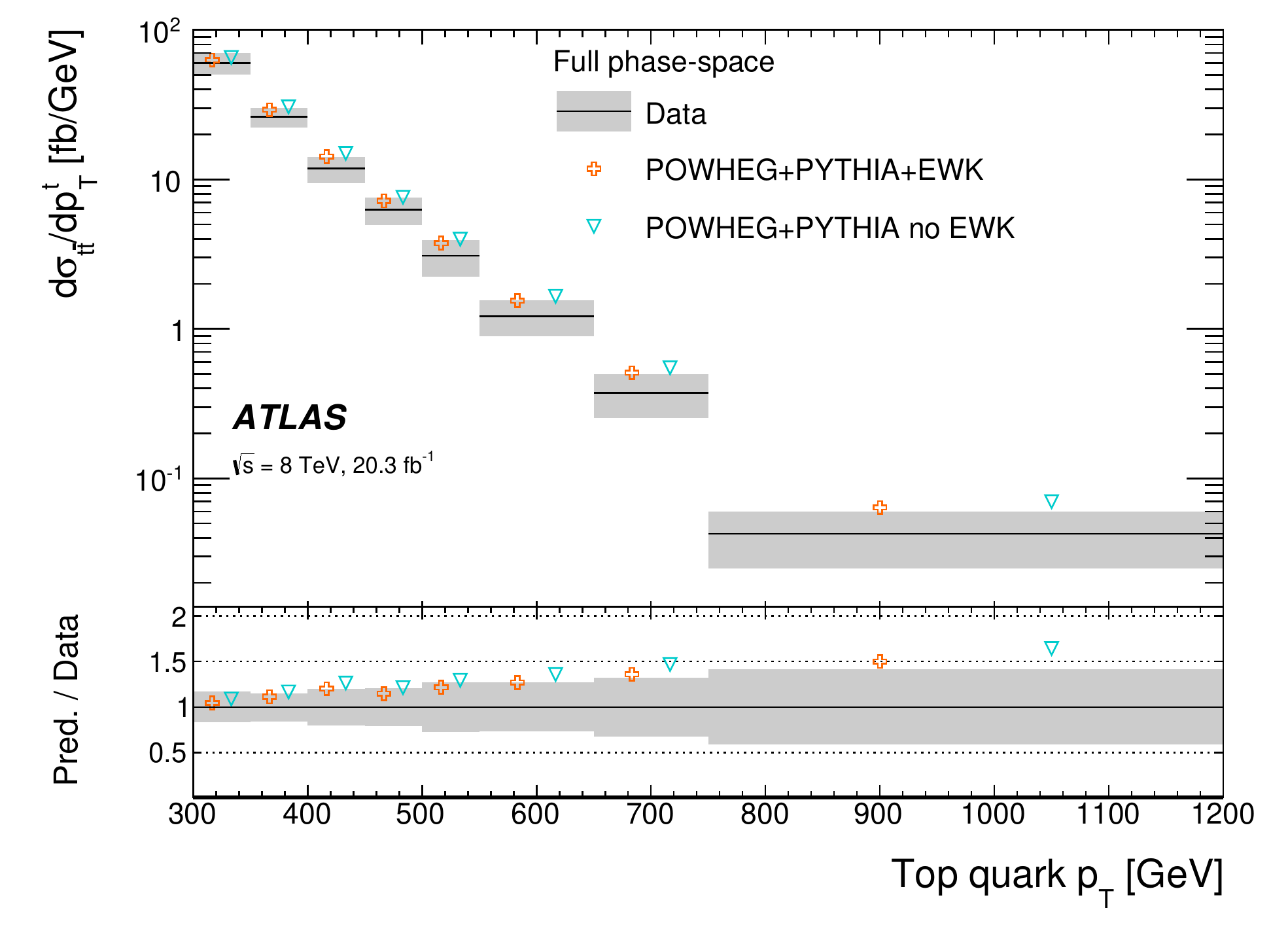}}

  \caption{(a) Fiducial particle-level differential cross-section as a
    function of the hadronic top-jet candidate $\pt$ and (b) parton-level differential cross-section as a
    function of the hadronically decaying top quark $\pt$, both compared
    to the
    {\sc Powheg+Pythia}  predictions with and without electroweak
    corrections applied. 
          MC samples are normalized to the NNLO+NNLL inclusive cross-section $\sigma_{t\bar{t}}= 253$~pb.
          The lower part of the figure shows the ratio of
          the MC prediction to the data. 
	  The shaded area includes the total statistical plus systematic uncertainties.  The points of the various predictions are spaced along
        the horizontal axis for presentation only; they
        correspond to the same $\pt$ range.}
  \label{fig:ewkcorr}
\end{figure*}

\begin{figure*}
\centering 
\subfigure[\hspace*{0.15cm} ]{\includegraphics[width=0.49\textwidth]{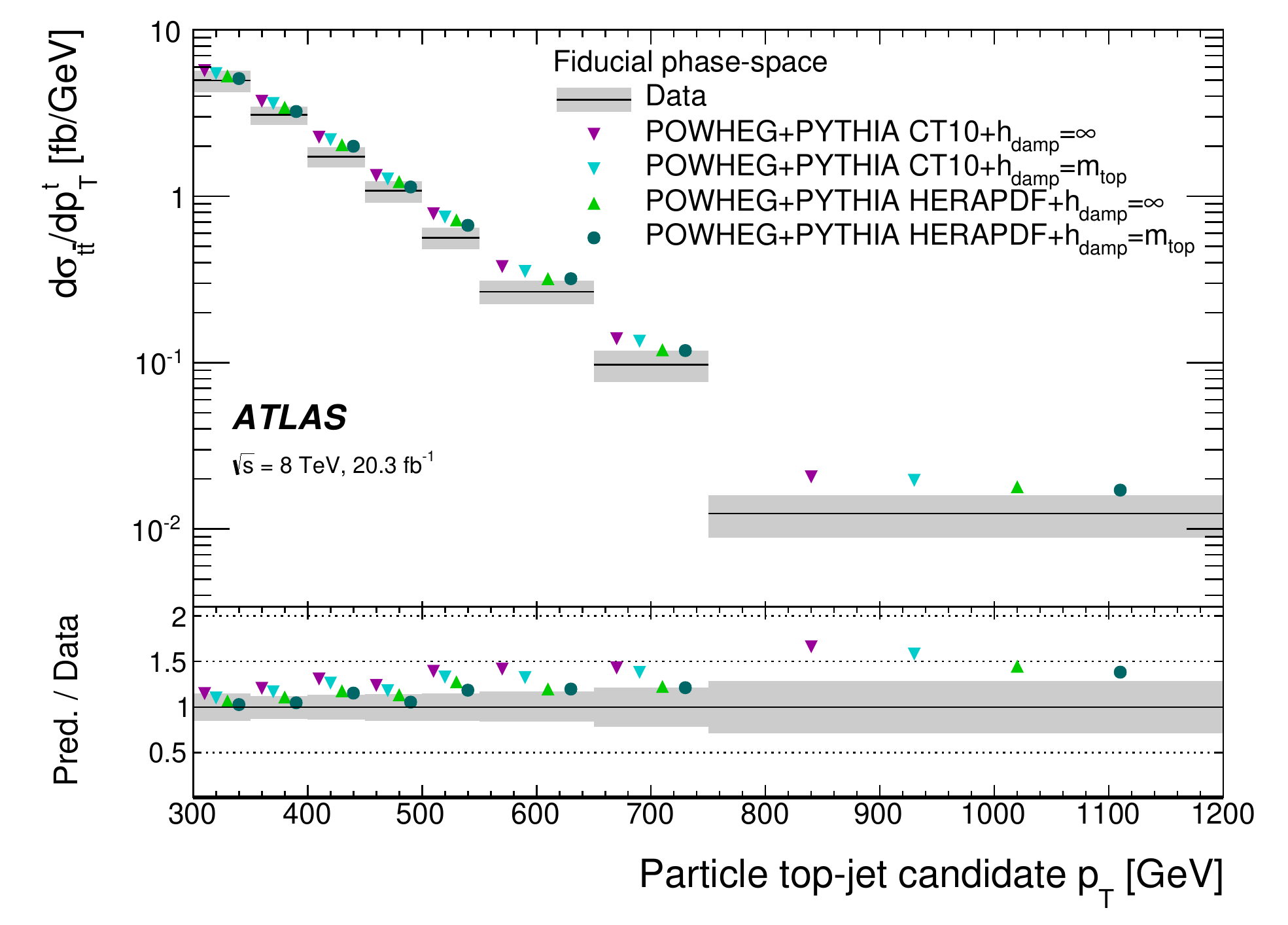}}
  \subfigure[\hspace*{0.15cm} ]{\includegraphics[width=0.49\textwidth]{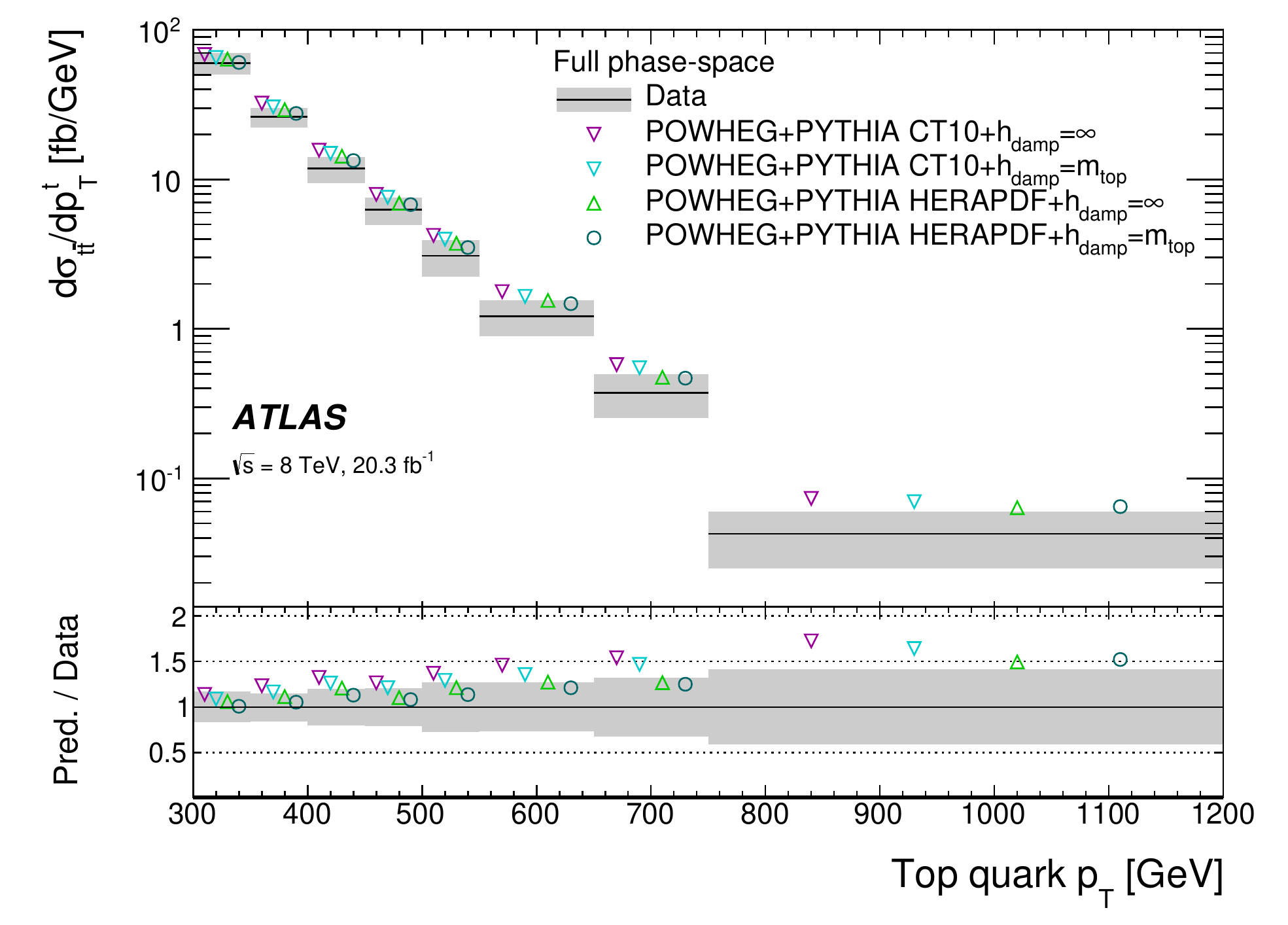}}
  \caption{(a) Fiducial particle-level differential cross-section as a
    function of the hadronic top-jet candidate $\pt$ and (b) parton-level differential cross-section as a
    function of the hadronically decaying top quark $\pt$, both compared
    to 
    {\sc Powheg+Pythia}  predictions using either the {\sc  HERAPDF} or {\sc
  CT10} PDF sets, and the {\sc Powheg} $h_\mathrm{damp}$ parameter set to
$\infty$ or $m_\mathrm{top}$. 
          MC samples are normalized to the NNLO+NNLL inclusive cross-section $\sigma_{t\bar{t}}= 253$~pb.
	  No electroweak corrections are applied to the predictions.
          The lower part of the figure shows the ratio of
          the MC prediction to the data.
	  The shaded area includes the total statistical plus systematic uncertainties.  The points of the various predictions are spaced along
        the horizontal axis for presentation only; they
        correspond to the same $\pt$ range.}
  \label{fig:hdamp_herapdf}
\end{figure*}

The measured differential cross-sections are  compared in Fig.~\ref{fig:ewkcorr} to
the predictions of {\sc Powheg}+{\sc Pythia} with and without the
electroweak corrections applied. The electroweak corrections
lead to a slightly softer $\pt$ spectrum, increasing the particle-level $p$-value from 0.11 to
0.28 without and with the corrections, respectively. 
The measured differential cross-sections  are also
compared in Fig.~\ref{fig:hdamp_herapdf} to {\sc Powheg}+{\sc Pythia}
predictions using either the {\sc  HERAPDF} \cite{Aaron:2009aa} or {\sc
  CT10} PDF sets, and two different values of the {\sc Powheg}
$h_\mathrm{damp}$ parameter, the nominal value $h_\mathrm{damp} =
m_\mathrm{top}$ and one with $h_\mathrm{damp} = \infty$,
which increases the amount of hard radiation and yields a lower
$p$-value of 0.05.
Better agreement with data is obtained when using the {\sc
  HERAPDF} set instead of {\sc CT10}, which reduces the difference
between data and MC simulation
by up to about 20\%. The  {\sc Powheg}+{\sc Pythia}
prediction that provides the best description of the data is the one that simultaneously
employs the {\sc  HERAPDF} set and  $h_\mathrm{damp} = m_\mathrm{top}$,
corresponding to a $p$-value of 0.31 at particle level.

\begin{figure*}[htbp]
\centering
\includegraphics[width=0.5\textwidth]{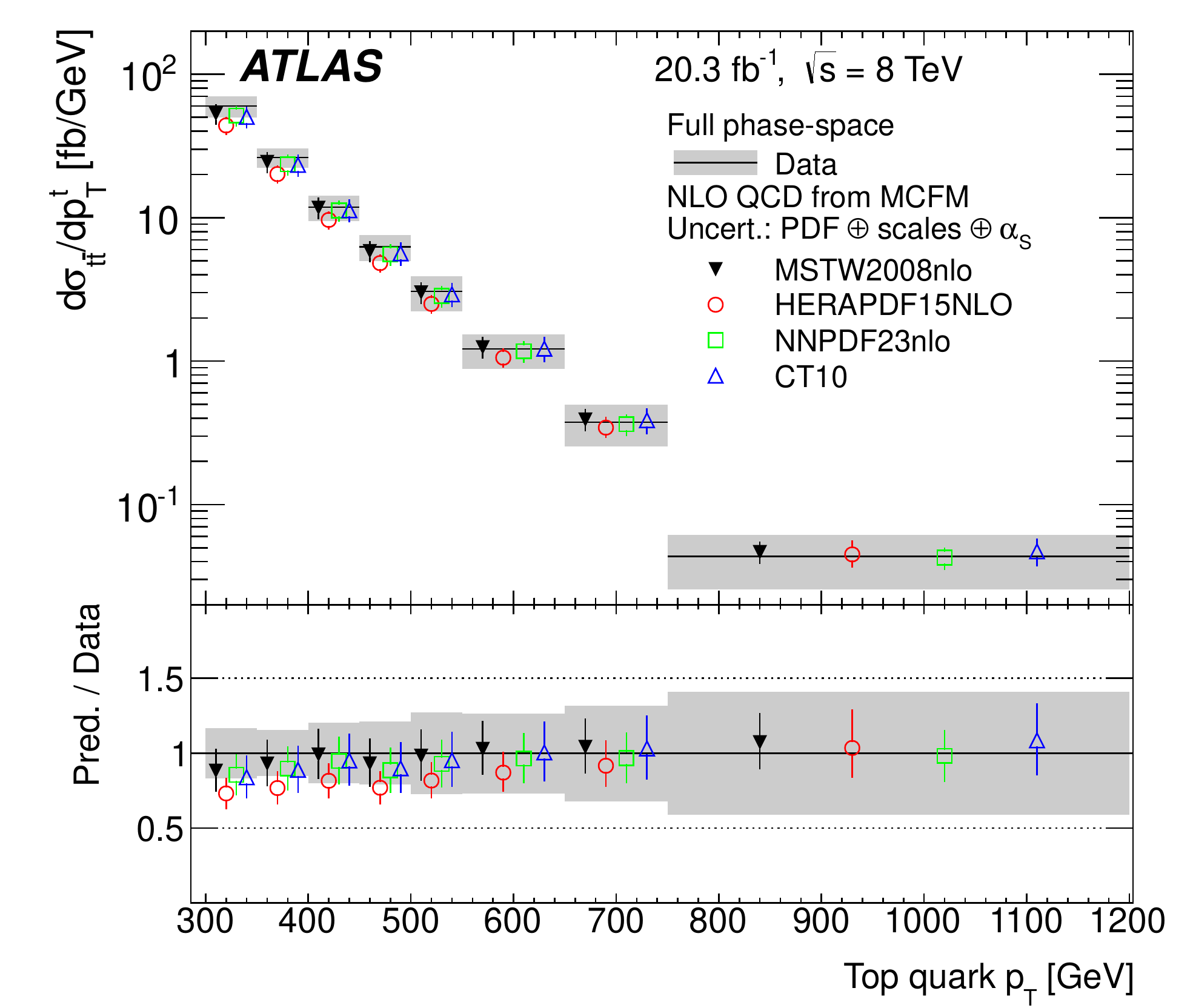}  
 \caption{Parton-level differential cross-section as a
    function of the hadronically decaying top quark $\pt$. 
    {\sc MCFM} predictions with various PDF sets are also shown.
          The lower part of the figure shows the ratio of
          the MC prediction to the data. 
	  The shaded area includes the total statistical plus
          systematic uncertainties.  
          The uncertainty on the predictions include the PDF uncertainties and variations of $\alphas$, $\mu_\mathrm{F}$, $\mu_\mathrm{R}$.}
  \label{fig:theory_vs_data_totalUnc}
\end{figure*}

The measured parton-level cross-section is compared to the prediction
of the parton-level NLO {\sc MCFM} generator
\cite{Campbell:2010ff}, which is interfaced
with Applgrid \cite{Carli:2010rw} to convolve the perturbative
coefficients with the strong coupling and the PDF. The inclusive
cross-section computed by {\sc
  MCFM} is used to normalize the prediction and 
no electroweak corrections are applied.
Several PDF
sets are compared: CT10, MSTW, NNPDF, and HERAPDF. The renormalization
scale $\mu_\mathrm{R}$ and 
factorization scale $\mu_\mathrm{F}$ are dynamic:
$\mu_\mathrm{R}=\mu_\mathrm{F}=\sqrt{m_\mathrm{top}^2+ \hat{p}_{\mathrm{T,top}}^2 }$,
where $\hat{p}_{\mathrm{T,top}}$ is the average $\pt$ of the
two top quarks in the event. 
The uncertainties on the prediction include the PDF uncertainties 
estimated according to the prescription of each set
and variations of the strong
coupling constant, $\mu_\mathrm{F}$, and $\mu_\mathrm{R}$. 
The predictions are compared
to the measured parton-level cross-section in
Fig.~\ref{fig:theory_vs_data_totalUnc}.
All predictions are in good
agreement with the measured cross-section within the quoted
uncertainties, which are dominated by systematic uncertainties
correlated between $\pt$ bins.

%% file: sum_pt_hadTop_SVD_particle__ResultTable.tex
\begin{tabular}{lccc} \hline
$\ptptcl$[\GeV] & $\frac{d\sigma_{t\bar{t}}}{d\ptptcl}$ $\left[ \frac{\rm fb}{\rm \GeV} \right]$ & Statistical [$\%$] & Systematic [$\%$] \\ \hline
300 -- 350 & 4.97 & $\pm$2.7 & $\pm$15 \\
350 -- 400 & 3.09 & $\pm$3.5 & $\pm$13 \\
400 -- 450 & 1.73 & $\pm$4.2 & $\pm$13 \\
450 -- 500 & 1.08 & $\pm$4.4 & $\pm$14 \\
500 -- 550 & 0.56 & $\pm$6.1 & $\pm$14 \\
550 -- 650 & 0.27 & $\pm$6.0 & $\pm$16 \\
650 -- 750 & 0.097 & $\pm$8.1 & $\pm$20 \\
750 -- 1200 & 0.012 & $\pm$15 & $\pm$24 \\
\hline
\end{tabular}

%% file: sum_pt_hadTop_SVD_parton__ResultTable.tex
\begin{tabular}{lccc} \hline
$\ptptn$ [\GeV] & $\frac{d\sigma_{t\bar{t}}}{d\ptptcl}$ $\left[ \frac{\rm fb}{\rm \GeV} \right]$ & Statistical [$\%$] & Systematic [$\%$] \\ 
\hline
300 -- 350 &  60.1    & $\pm$3.2 &  $\pm$16    \\ 
350 -- 400 &  26.2    & $\pm$3.4 &  $\pm$15    \\ 
400 -- 450 &  11.8    & $\pm$4.2 &  $\pm$20    \\ 
450 -- 500 &  6.27    & $\pm$4.5 &  $\pm$21    \\ 
500 -- 550 &  3.06    & $\pm$6.1 &  $\pm$27    \\ 
550 -- 650 &  1.21    & $\pm$6.3 &  $\pm$26    \\ 
650 -- 750 &  0.375   & $\pm$9.6 &  $\pm$31     \\ 
750 -- 1200 & 0.043   & $\pm$17  &  $\pm$38    \\
\hline
\end{tabular}

%% file: pvaluesTableCombined.tex
\begin{table*} [p]
\centering
\begin{tabular}{llrr} \hline
MC generator & PDF  & ~~~$\chi^{2}$~~~ &  ~~~$p$-value~~ \\ \hline 
\Powheg{}+\Pythia{} $h_\mathrm{damp}=m_\mathrm{top}$ + Electroweak corr. & CT10 & 9.8 & 0.28  \\
\Powheg{}+\Pythia{} $h_\mathrm{damp}=m_\mathrm{top}$ & CT10 & 13.0 & 0.11  \\
\Powheg{}+\Pythia{} $h_\mathrm{damp}=\infty$ & CT10 & 15.6 & 0.05  \\
\Powheg{}+\Pythia{} $h_\mathrm{damp}=m_\mathrm{top}$ & HERAPDF & 9.4 & 0.31  \\
\Powheg{}+\Pythia{} $h_\mathrm{damp}=\infty$ & HERAPDF & 10.9 & 0.21  \\
\Powheg{}+\Herwig{} & CT10 & 8.2 & 0.41  \\
\McAtNlo{}+\Herwig{} & CT10 & 12.3 & 0.14  \\
\Alpgen{}+\Herwig{} & CTEQ6 & 33.1 & $5.9\cdot 10^{-5}$  \\
\hline
\end{tabular}
\caption{Values of $\chi^2$ and a $p$-value, computed for 8 degrees of freedom, obtained from the
  covariance matrix of the measured cross-section for various
  predictions. 
  Electroweak corrections are applied only to the first prediction.
}
\label{tab:pvalues_combined}
\end{table*}

%% file: conclusions.tex
\section{Conclusions}
\label{sec:conclusions}

The differential $\ttbar$ production cross-section in \rts\ = 8 \TeV\ $pp$ collisions 
has been measured as a
function of the hadronically decaying top quark $\pt$ in a high-$\pt$ regime, using a
dataset corresponding to an integrated luminosity 
of 20.3 \ifb\ collected by the ATLAS detector at the LHC. 
Boosted hadronically decaying top quarks with $\pt > 300 $ \GeV\
are reconstructed 
within  large-$R$ jets and identified using jet substructure
techniques. The measured $\pt$
spectrum is extended in this analysis
relative to previous measurements. 
A particle-level cross-section is measured in a 
fiducial region that closely follows the event 
selection. The measurement uncertainty ranges from 13\%
to 29\% and is generally dominated by the uncertainty on the jet energy scale of large-$R$ jets.
A parton-level cross-section is also
reported, with larger systematic uncertainties due to its greater
reliance on $\ttbar$ MC generators to correct the data.
The measured cross-sections are compared to the predictions of several
NLO and LO matrix-element generators normalized to NNLO+NNLL QCD
calculations, and using various PDF sets. 
Previous measurements suggest that the top quark $\pt$ spectrum is well
predicted at low $\pt$ by NLO and matrix-element MC generators,
both in normalization and shape, but that their predictions exceed the
data at high $\pt$. The current analysis, focused on the boosted
topology and extended to higher $\pt$ values, also observes such a trend. However, a statistical analysis
shows that the measurements are
compatible with the majority of MC generator predictions within the quoted
uncertainties.  

%% file: Acknowledgements.tex

We thank CERN for the very successful operation of the LHC, as well as the
support staff from our institutions without whom ATLAS could not be
operated efficiently.

We acknowledge the support of ANPCyT, Argentina; YerPhI, Armenia; ARC, Australia; BMWFW and FWF, Austria; ANAS, Azerbaijan; SSTC, Belarus; CNPq and FAPESP, Brazil; NSERC, NRC and CFI, Canada; CERN; CONICYT, Chile; CAS, MOST and NSFC, China; COLCIENCIAS, Colombia; MSMT CR, MPO CR and VSC CR, Czech Republic; DNRF, DNSRC and Lundbeck Foundation, Denmark; IN2P3-CNRS, CEA-DSM/IRFU, France; GNSF, Georgia; BMBF, HGF, and MPG, Germany; GSRT, Greece; RGC, Hong Kong SAR, China; ISF, I-CORE and Benoziyo Center, Israel; INFN, Italy; MEXT and JSPS, Japan; CNRST, Morocco; FOM and NWO, Netherlands; RCN, Norway; MNiSW and NCN, Poland; FCT, Portugal; MNE/IFA, Romania; MES of Russia and NRC KI, Russian Federation; JINR; MESTD, Serbia; MSSR, Slovakia; ARRS and MIZ\v{S}, Slovenia; DST/NRF, South Africa; MINECO, Spain; SRC and Wallenberg Foundation, Sweden; SERI, SNSF and Cantons of Bern and Geneva, Switzerland; MOST, Taiwan; TAEK, Turkey; STFC, United Kingdom; DOE and NSF, United States of America. In addition, individual groups and members have received support from BCKDF, the Canada Council, CANARIE, CRC, Compute Canada, FQRNT, and the Ontario Innovation Trust, Canada; EPLANET, ERC, FP7, Horizon 2020 and Marie Sk\u0142odowska-Curie Actions, European Union; Investissements d'Avenir Labex and Idex, ANR, Region Auvergne and Fondation Partager le Savoir, France; DFG and AvH Foundation, Germany; Herakleitos, Thales and Aristeia programmes co-financed by EU-ESF and the Greek NSRF; BSF, GIF and Minerva, Israel; BRF, Norway; the Royal Society and Leverhulme Trust, United Kingdom.

The crucial computing support from all WLCG partners is acknowledged
gratefully, in particular from CERN and the ATLAS Tier-1 facilities at
TRIUMF (Canada), NDGF (Denmark, Norway, Sweden), CC-IN2P3 (France),
KIT/GridKA (Germany), INFN-CNAF (Italy), NL-T1 (Netherlands), PIC (Spain),
ASGC (Taiwan), RAL (UK) and BNL (USA) and in the Tier-2 facilities
worldwide.

%% file: systBreakDown.tex
\section{Detailed tables of systematic uncertainties}
\label{Sec:Appendix:Syst}

Tables \ref{tab:CombSyst_ptptcl} and \ref{tab:CombSyst_ptgen} report the detailed 
breakdown of the systematic uncertainties as a percentage of the measured differential 
cross sections.

\begin{table*} [htbp]
\begin{center}
\input{sum_pt_hadTop_SVD_particle__SysTable.tex}
\caption{The individual systematic uncertainties calculated as
  a~percentage of the differential cross-section $\dsigma_{\ttbar}/d\ptptcl$ in each bin.}
\label{tab:CombSyst_ptptcl}
\end{center}
\end{table*}
\begin{table*} [htbp]
\begin{center}
\input{sum_pt_hadTop_SVD_parton__SysTable.tex}
\caption{The individual systematic uncertainties calculated as
  a~percentage of the differential cross-section $\dsigma_{\ttbar}/d\ptptn$ in each bin.}
\label{tab:CombSyst_ptgen}
\end{center}
\end{table*}

\clearpage

%% file: sum_pt_hadTop_SVD_particle__SysTable.tex
{\tiny
\noindent\makebox[\textwidth]{
\begin{tabular}{l c c c c c c c c}
\hline
\hline
$\dsigma_{\ttbar}/d\ptptcl$ Uncertainties [\%] $/$ Bins [\GeV] &     300--350 &     350--400 &     400--450 &     450--500 &     500--550 &  550--650 &     650--750 & 750--1200 \\ 
\hline
\hline
Large-$R$ jet $p_{\mathrm{T}}$ resolution & 3.9/$-$4.0 & $-$3.9/3.9 & 2.6/$-$2.6 & 1.3/$-$1.3 &  - / -  & 0.7/$-$0.7 & 2.6/$-$2.6 & 1.6/$-$1.5\\ [0.5mm]
Large-$R$ jet mass resolution & $-$0.5/0.5 & $-$0.2/0.2 & $-$0.2/0.2 & $-$0.2/0.2 & $-$0.3/0.3 & $-$0.3/0.3 & $-$0.7/0.7 & $-$0.7/0.7\\ [0.5mm]
Large-$R$ jet $\sqrt{d_{12}}$ scale & 1.0/$-$1.0 & 1.1/$-$1.0 & 0.8/$-$1.1 & 0.8/$-$1.3 & 0.9/$-$1.3 & 1.0/$-$1.4 & 1.4/$-$1.8 & 1.8/$-$2.5\\ [0.5mm]
Large-$R$ jet mass scale & 4.0/$-$4.5 & 2.5/$-$2.5 & 2.1/$-$2.0 & 1.7/$-$2.1 & 1.4/$-$1.6 & 1.3/$-$1.4 & 1.7/$-$2.2 & 2.2/$-$3.0\\ [0.5mm]
Large-$R$ jet (JES) data vs MC & 1.6/$-$2.3 & 4.7/$-$4.6 & 5.5/$-$5.7 & 6.4/$-$6.1 & 6.5/$-$6.0 & 7.3/$-$7.1 & 10.0/$-$9.6 & 11.7/$-$11.4\\ [0.5mm]
Large-$R$ jet (JES) validation of $\Delta\phi$ cut &  - / -  &  - / -  & 0.1/$-$0.2 & 0.3/ -  &  - /$-$0.2 &  - /$-$0.3 & 0.5/$-$0.4 & 0.8/$-$0.5\\ [0.5mm]
Large-$R$ jet (JES) cut on subleading small-$R$ jet & 0.9/$-$0.8 & 0.5/$-$1.0 & 1.2/$-$0.9 & 1.3/$-$1.0 & 1.3/$-$1.6 & 1.9/$-$2.7 & 2.8/$-$2.8 & 2.8/$-$2.9\\ [0.5mm]
Large-$R$ jet (JES) photon purity & 0.2/ -  &  - / -  &  - / -  &  - / -  & $-$0.1/ -  & $-$0.2/ -  &  - / -  &  - /$-$0.4\\ [0.5mm]
Large-$R$ jet (JES) photon energy scale & 1.0/$-$0.9 & 1.7/$-$2.0 & 2.6/$-$2.4 & 2.9/$-$2.8 & 3.0/$-$3.2 & 3.0/$-$3.7 & 4.4/$-$3.9 & 5.6/$-$4.4\\ [0.5mm]
Large-$R$ jet (JES) generator & 0.8/$-$0.9 & 1.0/$-$1.1 & 1.3/$-$1.2 & 1.3/$-$0.8 & 0.5/$-$1.1 & 0.9/$-$1.6 & 1.5/$-$1.2 & 1.6/$-$1.2\\ [0.5mm]
Large-$R$ jet (JES) out of cone and underlying events & 0.2/$-$0.2 & 0.2/ -  &  - /$-$0.3 & 0.2/ -  &  - /$-$0.4 &  - /$-$0.6 & 0.5/$-$0.4 & 0.1/$-$0.4\\ [0.5mm]
Large-$R$ jet (JES) JER & 0.1/ -  &  - / -  &  - / -  &  - / -  &  - /$-$0.2 &  - /$-$0.2 & 0.5/$-$0.4 & 0.4/$-$0.9\\ [0.5mm]
Large-$R$ jet (JES) definition of small-$R$ jet inside large-$R$ jet &  - /0.2 & 0.6/$-$1.0 & 1.5/$-$1.4 & 1.7/$-$1.2 & 1.3/$-$1.5 & 1.3/$-$2.3 & 2.2/$-$2.3 & 2.9/$-$2.5\\ [0.5mm]
Large-$R$ jet (JES) cut on leading small-$R$ jet & 0.2/$-$0.2 & 0.4/$-$0.3 & 0.3/$-$0.3 &  - /$-$0.1 &  - / -  &  - /$-$0.2 & 0.1/$-$0.4 &  - /$-$0.7\\ [0.5mm]
Large-$R$ jet (JES) statistics & 0.3/$-$0.1 &  - /$-$0.7 & 1.1/$-$0.6 & 1.9/$-$2.0 & 2.1/$-$2.6 & 4.0/$-$4.3 & 8.0/$-$7.9 & 10.9/$-$10.7\\ [0.5mm]
Large-$R$ jet (JES) correlation with JMS & 1.1/$-$0.9 & 1.8/$-$2.1 & 2.6/$-$2.0 & 2.9/$-$2.7 & 2.2/$-$3.3 & 2.9/$-$3.5 & 4.0/$-$3.3 & 4.2/$-$3.8\\ [0.5mm]
Large-$R$ jet (JES) interpolation &  - / -  &  - / -  &  - / -  &  - / -  & $-$0.1/ -  & $-$0.5/0.2 & $-$0.7/0.6 & $-$0.6/ - \\ [0.5mm]
Large-$R$ jet (JES) topology & 11.3/$-$11.3 & 7.5/$-$5.9 & 7.8/$-$7.9 & 9.4/$-$8.3 & 8.1/$-$7.6 & 6.0/$-$5.9 & 7.7/$-$7.6 & 8.9/$-$8.7\\ [0.5mm]
Large-$R$ jet (JES) pileup offset $\mu$ & $-$0.3/0.3 & $-$0.2/0.2 & $-$0.8/0.6 & $-$0.3/0.2 & $-$0.2/0.2 & $-$0.8/0.4 & $-$1.3/1.0 & $-$1.1/1.6\\ [0.5mm]
Large-$R$ jet (JES) pileup offset $N_\mathrm{PV}$ &  - /0.2 & $-$0.1/ -  & $-$0.2/0.1 & $-$0.2/0.4 & $-$0.5/ -  & $-$0.4/ -  & $-$0.5/0.5 & $-$0.4/0.2\\ [0.5mm]
Small-$R$ jet JES & 0.4/$-$0.7 & 0.8/$-$1.3 & 1.5/$-$1.8 & 1.8/$-$1.6 & 1.7/$-$1.9 & 1.8/$-$3.0 & 2.3/$-$2.8 & 3.1/$-$3.1\\ [0.5mm]
Small-$R$ jet reconstruction efficiency &  - / -  &  - / -  &  - / -  &  - / -  &  - / -  & $-$0.1/0.1 &  - / -  &  - / - \\ [0.5mm]
Small-$R$ jet energy resolution & $-$0.2/0.2 & $-$0.8/0.8 &  - / -  & $-$0.7/0.7 & $-$1.3/1.3 & $-$0.8/0.7 &  - /$-$0.1 & $-$1.5/1.4\\ [0.5mm]
Small-$R$ jet JVF &  - /0.2 &  - /0.4 &  - /0.2 &  - /0.2 &  - /0.2 &  - /0.2 &  - /0.5 &  - /0.5\\ [0.5mm]
$b$-tagging $b$-jet efficiency & 1.4/$-$1.1 & 1.6/$-$1.4 & 2.6/$-$2.5 & 3.5/$-$3.4 & 3.6/$-$3.4 & 4.6/$-$4.7 & 5.8/$-$6.7 & 5.6/$-$6.9\\ [0.5mm]
$b$-tagging $c$-jet efficiency & 0.6/$-$0.6 & 0.6/$-$0.6 &  - / -  & $-$0.2/0.2 & $-$0.8/0.8 & $-$1.6/1.6 & $-$2.3/2.2 & $-$1.9/1.9\\ [0.5mm]
$b$-tagging light-jet efficiency & 0.3/$-$0.3 & 0.3/$-$0.3 & 0.4/$-$0.4 & 0.6/$-$0.6 & 0.7/$-$0.6 & 0.8/$-$0.8 & $-$1.0/0.8 & $-$4.7/4.0\\ [0.5mm]
$e$ efficiency & 0.6/$-$0.6 & 0.6/$-$0.6 & 0.6/$-$0.6 & 0.6/$-$0.6 & 0.6/$-$0.6 & 0.7/$-$0.7 & 0.6/$-$0.6 & 0.6/$-$0.6\\ [0.5mm]
$e$ energy resolution & $-$0.2/ -  &  - / -  & $-$0.2/ -  &  - / -  &  - /0.2 & $-$0.4/ -  &  - / -  &  - / - \\ [0.5mm]
$e$ energy scale & $-$0.7/0.3 & $-$0.9/0.6 & $-$1.1/0.6 & $-$1.2/0.8 & $-$1.3/1.1 & $-$1.3/0.7 & $-$1.0/0.9 & $-$0.9/1.1\\ [0.5mm]
$\mu$ efficiency & 0.9/$-$0.9 & 0.9/$-$0.9 & 0.9/$-$1.0 & 0.8/$-$1.0 & 0.9/$-$1.0 & 1.1/$-$0.9 & 1.0/$-$0.8 & 1.2/$-$0.9\\ [0.5mm]
$\mu$ ID momentum resolution &  - / -  &  - / -  &  - / -  & $-$0.1/ -  &  - / -  &  - /0.2 &  - / -  & $-$0.2/ - \\ [0.5mm]
$\mu$ MS momentum resolution &  - / -  &  - / -  &  - / -  &  - / -  &  - / -  & 0.2/ -  & 0.1/ -  &  - / - \\ [0.5mm]
$\mu$ momentum scale &  - / -  &  - / -  &  - / -  &  - / -  &  - /$-$0.1 &  - / -  &  - /$-$0.1 &  - /$-$0.3\\ [0.5mm]
$\met$ unassociated cells resolution &  - / -  & 0.1/ -  & 0.1/ -  & $-$0.2/ -  & $-$0.2/ -  &  - / -  &  - / -  &  - /$-$0.3\\ [0.5mm]
$\met$ unassociated cells scale & 0.2/ -  &  - / -  &  - /$-$0.1 &  - /0.1 &  - / -  & $-$0.2/ -  &  - / -  & $-$0.2/ - \\ [0.5mm]
Luminosity & 2.8/$-$2.8 & 2.8/$-$2.8 & 2.9/$-$2.9 & 2.9/$-$2.9 & 2.9/$-$2.9 & 2.9/$-$2.9 & 2.9/$-$2.9 & 2.9/$-$2.9\\ [0.5mm]
$W$+jet & 0.4/$-$0.4 & 0.2/$-$0.2 & 0.2/$-$0.2 & 0.2/$-$0.2 & 0.4/$-$0.4 & 0.9/$-$0.9 & 2.7/$-$2.7 & 5.4/$-$5.5\\ [0.5mm]
Single top & 1.3/$-$1.3 & 0.6/$-$0.6 & 1.6/$-$1.6 & 1.5/$-$1.5 & 2.8/$-$2.8 & 4.4/$-$4.4 & 4.4/$-$4.4 & 3.9/$-$3.9\\ [0.5mm]
$Z$+jets & 0.3/$-$0.3 & 0.2/$-$0.2 & 0.2/$-$0.2 & 0.4/$-$0.4 & 0.5/$-$0.5 & 0.7/$-$0.7 & 0.4/$-$0.4 & 0.3/$-$0.3\\ [0.5mm]
Multijet & $-$0.1/ -  & $-$0.1/0.1 &  - / -  & 0.2/$-$0.2 &  - / -  & $-$0.3/0.3 & $-$0.1/0.1 & $-$0.1/0.1\\ [0.5mm]
Diboson  & 0.2/$-$0.2 & 0.2/$-$0.2 & 0.4/$-$0.4 & 0.2/$-$0.2 & 0.3/$-$0.3 & 0.4/$-$0.4 & 0.6/$-$0.6 & 0.7/$-$0.7\\ [0.5mm]
MC Signal statistics & 0.5/$-$0.5 & 0.7/$-$0.7 & 0.9/$-$0.9 & 0.9/$-$0.9 & 1.3/$-$1.3 & 1.0/$-$1.0 & 1.1/$-$1.1 & 1.7/$-$1.7\\ [0.5mm]
MC Background statistics & 0.5/$-$0.5 & 0.5/$-$0.5 & 0.7/$-$0.7 & 0.8/$-$0.8 & 1.2/$-$1.2 & 1.4/$-$1.4 & 1.8/$-$1.8 & 3.1/$-$3.1\\ [0.5mm]
$t\overline{t}$ generator & 3.4/$-$3.4 & 3.6/$-$3.6 & 0.6/$-$0.6 & 0.4/$-$0.4 &  - / -  & 3.1/$-$3.1 & 1.0/$-$1.0 & 5.3/$-$5.3\\ [0.5mm]
PS/Hadronization & 1.6/$-$1.6 & 2.1/$-$2.1 & 3.6/$-$3.6 & 0.3/$-$0.3 & 1.1/$-$1.1 & 2.5/$-$2.5 & 2.2/$-$2.2 & 0.4/$-$0.4\\ [0.5mm]
ISR/FSR & $-$4.0/4.0 & $-$4.0/4.0 & $-$3.6/3.6 & $-$3.6/3.6 & $-$3.6/3.6 & $-$3.6/3.6 & $-$5.5/5.5 & $-$5.5/5.5\\ [0.5mm]
PDF &  - / -  &  - / -  &  - / -  & 0.8/$-$0.8 & 0.8/$-$0.8 & 1.2/$-$1.2 & 1.2/$-$1.2 & 1.2/$-$1.2\\ [0.5mm]
\hline
\end{tabular}}
}

%% file: sum_pt_hadTop_SVD_parton__SysTable.tex
{\tiny
\noindent\makebox[\textwidth]{
\begin{tabular}{l c c c c c c c c}
\hline
\hline
$\dsigma_{\ttbar}/d\ptptn$ Uncertainties [\%] $/$ Bins [\GeV] &     300--350 &     350--400 &     400--450 &     450--500 &     500--550 &  550--650 &     650--750 & 750--1200 \\ 
\hline
\hline
Large-$R$ jet $p_{\mathrm{T}}$ resolution & 4.7/$-$4.8 & $-$4.5/4.1 & 1.6/$-$1.7 & 2.0/$-$2.0 & 0.4/$-$0.3 & 0.7/$-$0.7 & 2.0/$-$2.0 & 2.3/$-$2.2\\ [0.5mm]
Large-$R$ jet mass resolution & $-$0.5/0.5 & $-$0.2/0.2 & $-$0.2/0.2 & $-$0.2/0.2 & $-$0.2/0.2 & $-$0.4/0.4 & $-$0.7/0.6 & $-$0.8/0.8\\ [0.5mm]
Large-$R$ jet $\sqrt{d_{12}}$ scale & 1.0/$-$1.0 & 1.1/$-$1.0 & 0.8/$-$1.1 & 0.8/$-$1.2 & 0.8/$-$1.3 & 1.0/$-$1.5 & 1.5/$-$2.0 & 1.9/$-$2.6\\ [0.5mm]
Large-$R$ jet mass scale & 4.3/$-$4.8 & 2.5/$-$2.5 & 1.9/$-$1.8 & 1.6/$-$1.9 & 1.2/$-$1.5 & 1.2/$-$1.4 & 1.6/$-$2.1 & 2.0/$-$2.8\\ [0.5mm]
Large-$R$ jet (JES) data vs MC & 1.1/$-$1.9 & 4.7/$-$4.6 & 5.9/$-$6.0 & 6.5/$-$6.2 & 6.8/$-$6.3 & 8.0/$-$7.6 & 10.8/$-$10.4 & 12.8/$-$12.5\\ [0.5mm]
Large-$R$ jet (JES) validation of $\Delta\phi$ cut &  - / -  &  - / -  & 0.1/$-$0.2 & 0.2/ -  &  - /$-$0.1 &  - /$-$0.3 & 0.6/$-$0.4 & 0.9/$-$0.6\\ [0.5mm]
Large-$R$ jet (JES) cut on subleading small-$R$ jet & 0.9/$-$0.8 & 0.5/$-$1.0 & 1.1/$-$0.9 & 1.3/$-$1.0 & 1.4/$-$1.7 & 2.1/$-$2.8 & 2.9/$-$3.2 & 3.2/$-$3.3\\ [0.5mm]
Large-$R$ jet (JES) photon purity & 0.2/ -  &  - / -  &  - / -  &  - / -  & $-$0.1/ -  & $-$0.2/ -  &  - /$-$0.2 &  - /$-$0.3\\ [0.5mm]
Large-$R$ jet (JES) photon energy scale & 0.9/$-$0.7 & 1.6/$-$2.0 & 2.7/$-$2.5 & 3.0/$-$2.9 & 3.0/$-$3.4 & 3.4/$-$3.9 & 4.8/$-$4.3 & 6.0/$-$4.7\\ [0.5mm]
Large-$R$ jet (JES) generator & 0.8/$-$0.9 & 1.0/$-$1.1 & 1.4/$-$1.2 & 1.2/$-$0.9 & 0.7/$-$1.1 & 0.9/$-$1.5 & 1.4/$-$1.4 & 1.8/$-$1.2\\ [0.5mm]
Large-$R$ jet (JES) out of cone and underlying events & 0.2/$-$0.2 & 0.2/ -  & 0.1/$-$0.3 & 0.2/ -  &  - /$-$0.3 &  - /$-$0.6 & 0.3/$-$0.5 & 0.3/$-$0.5\\ [0.5mm]
Large-$R$ jet (JES) JER & 0.1/ -  &  - / -  &  - / -  &  - / -  &  - /$-$0.1 & 0.1/$-$0.3 & 0.5/$-$0.6 & 0.6/$-$0.9\\ [0.5mm]
Large-$R$ jet (JES) definition of small-$R$ jet inside large-$R$ jet & $-$0.2/0.4 & 0.5/$-$1.0 & 1.6/$-$1.6 & 1.8/$-$1.3 & 1.4/$-$1.6 & 1.5/$-$2.3 & 2.4/$-$2.6 & 3.2/$-$2.8\\ [0.5mm]
Large-$R$ jet (JES) cut on leading small-$R$ jet & 0.2/$-$0.2 & 0.4/$-$0.3 & 0.3/$-$0.3 &  - /$-$0.1 &  - / -  &  - /$-$0.2 &  - /$-$0.5 &  - /$-$0.7\\ [0.5mm]
Large-$R$ jet (JES) statistics & 0.3/ -  &  - /$-$0.6 & 1.1/$-$0.6 & 1.9/$-$1.8 & 2.4/$-$2.9 & 4.8/$-$5.2 & 9.1/$-$9.2 & 12.3/$-$12.3\\ [0.5mm]
Large-$R$ jet (JES) correlation with JMS & 0.9/$-$0.8 & 1.7/$-$2.1 & 2.7/$-$2.2 & 2.9/$-$2.7 & 2.5/$-$3.4 & 3.0/$-$3.7 & 4.1/$-$3.7 & 4.7/$-$3.9\\ [0.5mm]
Large-$R$ jet (JES) interpolation &  - / -  &  - / -  &  - / -  &  - /$-$0.1 & $-$0.2/ -  & $-$0.5/0.2 & $-$0.8/0.4 & $-$0.8/0.2\\ [0.5mm]
Large-$R$ jet (JES) topology & 11.8/$-$12.0 & 7.4/$-$5.7 & 7.4/$-$7.3 & 9.2/$-$8.4 & 8.0/$-$7.5 & 6.3/$-$6.2 & 7.3/$-$7.2 & 8.5/$-$8.4\\ [0.5mm]
Large-$R$ jet (JES) pileup offset $\mu$ & $-$0.3/0.3 & $-$0.2/0.2 & $-$0.8/0.6 & $-$0.4/0.3 & $-$0.3/0.1 & $-$0.8/0.4 & $-$1.3/1.2 & $-$1.4/1.8\\ [0.5mm]
Large-$R$ jet (JES) pileup offset $N_\mathrm{PV}$ &  - /0.2 & $-$0.1/ -  & $-$0.2/0.1 & $-$0.2/0.3 & $-$0.5/ -  & $-$0.5/ -  & $-$0.5/0.3 & $-$0.5/0.4\\ [0.5mm]
Small-$R$ jet JES & 0.6/$-$0.8 & 0.8/$-$1.3 & 1.6/$-$1.9 & 2.0/$-$1.7 & 1.8/$-$2.0 & 1.9/$-$3.0 & 2.5/$-$3.3 & 3.3/$-$3.5\\ [0.5mm]
Small-$R$ jet reconstruction efficiency &  - / -  &  - / -  &  - / -  &  - / -  &  - / -  & $-$0.1/0.1 &  - / -  &  - / - \\ [0.5mm]
Small-$R$ jet energy resolution & $-$0.1/0.1 & $-$0.8/0.8 & $-$0.1/0.1 & $-$0.6/0.6 & $-$1.2/1.2 & $-$0.9/0.8 & $-$0.6/0.5 & $-$1.1/1.0\\ [0.5mm]
Small-$R$ jet JVF & 0.2/ -  &  - /0.4 &  - /0.3 &  - /0.2 &  - /0.2 &  - /0.3 &  - /0.5 &  - /0.6\\ [0.5mm]
$b$-tagging $b$-jet efficiency & 1.3/$-$1.0 & 1.5/$-$1.3 & 2.6/$-$2.5 & 3.6/$-$3.5 & 4.0/$-$3.8 & 5.0/$-$5.2 & 6.1/$-$7.1 & 6.5/$-$8.0\\ [0.5mm]
$b$-tagging $c$-jet efficiency & 0.7/$-$0.7 & 0.6/$-$0.6 &  - / -  & $-$0.3/0.3 & $-$1.0/1.0 & $-$1.9/1.9 & $-$2.5/2.4 & $-$2.6/2.5\\ [0.5mm]
$b$-tagging light-jet efficiency & 0.3/$-$0.3 & 0.2/$-$0.2 & 0.4/$-$0.4 & 0.6/$-$0.6 & 0.9/$-$0.8 & 0.4/$-$0.5 & $-$2.0/1.6 & $-$4.9/4.0\\ [0.5mm]
$e$ efficiency & 0.6/$-$0.6 & 0.6/$-$0.6 & 0.6/$-$0.6 & 0.6/$-$0.6 & 0.7/$-$0.7 & 0.7/$-$0.7 & 0.6/$-$0.6 & 0.6/$-$0.6\\ [0.5mm]
$e$ energy resolution & $-$0.2/ -  &  - / -  & $-$0.1/ -  & $-$0.1/ -  & $-$0.1/0.1 & $-$0.3/ -  & $-$0.1/ -  &  - / - \\ [0.5mm]
$e$ energy scale & $-$0.6/0.3 & $-$0.9/0.6 & $-$1.1/0.6 & $-$1.2/0.8 & $-$1.3/1.0 & $-$1.3/0.9 & $-$1.0/0.9 & $-$0.9/1.1\\ [0.5mm]
$\mu$ efficiency & 0.9/$-$0.9 & 0.9/$-$0.9 & 0.9/$-$1.0 & 0.9/$-$1.0 & 0.9/$-$1.0 & 1.1/$-$0.9 & 1.1/$-$0.8 & 1.2/$-$0.8\\ [0.5mm]
$\mu$ ID momentum resolution &  - / -  &  - / -  &  - / -  & $-$0.1/ -  &  - / -  &  - /0.1 &  - / -  & $-$0.2/ - \\ [0.5mm]
$\mu$ MS momentum resolution &  - / -  &  - / -  &  - / -  &  - / -  &  - / -  & 0.2/ -  & 0.1/ -  &  - / - \\ [0.5mm]
$\mu$ momentum scale &  - / -  &  - / -  &  - / -  &  - / -  &  - / -  &  - / -  &  - /$-$0.2 &  - /$-$0.3\\ [0.5mm]
$\met$ unassociated cells resolution &  - / -  & 0.1/ -  & 0.2/ -  & $-$0.2/ -  & $-$0.2/ -  &  - / -  &  - / -  &  - /$-$0.3\\ [0.5mm]
$\met$ unassociated cells scale & 0.3/ -  &  - / -  &  - /$-$0.1 &  - / -  &  - / -  & $-$0.2/ -  & $-$0.2/ -  & $-$0.2/ - \\ [0.5mm]
Luminosity & 2.9/$-$2.9 & 2.8/$-$2.8 & 2.9/$-$2.9 & 2.9/$-$2.9 & 2.9/$-$2.9 & 2.9/$-$2.9 & 2.9/$-$2.9 & 2.9/$-$2.9\\ [0.5mm]
$W$+jet & 0.4/$-$0.4 & 0.2/$-$0.2 & 0.2/$-$0.2 & 0.2/$-$0.2 & 0.4/$-$0.3 & 1.3/$-$1.3 & 3.6/$-$3.7 & 5.8/$-$6.1\\ [0.5mm]
Single top & 1.4/$-$1.4 & 0.6/$-$0.6 & 1.4/$-$1.4 & 1.6/$-$1.6 & 3.0/$-$3.0 & 4.6/$-$4.6 & 5.0/$-$5.0 & 4.7/$-$4.7\\ [0.5mm]
$Z$+jets & 0.3/$-$0.3 & 0.2/$-$0.2 & 0.2/$-$0.2 & 0.4/$-$0.4 & 0.5/$-$0.5 & 0.7/$-$0.7 & 0.5/$-$0.5 & 0.3/$-$0.3\\ [0.5mm]
Multijet & $-$0.1/0.1 & $-$0.1/0.1 &  - / -  & 0.2/$-$0.2 &  - / -  & $-$0.3/0.2 & $-$0.2/0.2 & $-$0.2/0.2\\ [0.5mm]
Diboson  & 0.2/$-$0.2 & 0.2/$-$0.2 & 0.4/$-$0.4 & 0.3/$-$0.3 & 0.3/$-$0.3 & 0.5/$-$0.5 & 0.7/$-$0.7 & 0.8/$-$0.8\\ [0.5mm]
MC Signal statistics & 0.8/$-$0.8 & 0.9/$-$0.9 & 1.2/$-$1.2 & 1.1/$-$1.1 & 1.7/$-$1.7 & 1.4/$-$1.4 & 1.5/$-$1.5 & 2.9/$-$2.9\\ [0.5mm]
MC Background statistics & 0.6/$-$0.6 & 0.6/$-$0.6 & 0.7/$-$0.7 & 0.8/$-$0.8 & 1.1/$-$1.1 & 1.4/$-$1.4 & 2.0/$-$2.0 & 3.1/$-$3.1\\ [0.5mm]
$t\overline{t}$ generator & 2.0/$-$2.0 & 2.3/$-$2.3 & 4.2/$-$4.2 & 4.7/$-$4.7 & 10.6/$-$10.6 & 11.5/$-$11.5 & 14.8/$-$14.8 & 20.4/$-$20.4\\ [0.5mm]
PS/Hadronization & 6.2/$-$6.2 & 9.3/$-$9.3 & 14.1/$-$14.1 & 13.6/$-$13.6 & 19.6/$-$19.6 & 16.7/$-$16.7 & 16.1/$-$16.1 & 17.3/$-$17.3\\ [0.5mm]
ISR/FSR & $-$4.1/4.1 & $-$4.1/4.1 & $-$5.6/5.6 & $-$5.6/5.6 & $-$5.6/5.6 & $-$5.6/5.6 & $-$6.5/6.5 & $-$6.5/6.5\\ [0.5mm]
PDF & 0.3/$-$0.3 & 0.3/$-$0.3 & 0.3/$-$0.3 & 1.2/$-$1.2 & 1.2/$-$1.2 & 2.2/$-$2.2 & 2.2/$-$2.2 & 2.2/$-$2.2\\ [0.5mm]
\hline
\end{tabular}}
}

%% file: atlas_authlist.tex
\begin{flushleft}
{\Large The ATLAS Collaboration}

\bigskip

G.~Aad$^\textrm{\scriptsize 85}$,
B.~Abbott$^\textrm{\scriptsize 113}$,
J.~Abdallah$^\textrm{\scriptsize 151}$,
O.~Abdinov$^\textrm{\scriptsize 11}$,
R.~Aben$^\textrm{\scriptsize 107}$,
M.~Abolins$^\textrm{\scriptsize 90}$,
O.S.~AbouZeid$^\textrm{\scriptsize 158}$,
H.~Abramowicz$^\textrm{\scriptsize 153}$,
H.~Abreu$^\textrm{\scriptsize 152}$,
R.~Abreu$^\textrm{\scriptsize 116}$
% ...
\bigskip
\\
% ...
$^{11}$ Institute of Physics, Azerbaijan Academy of Sciences, Baku, Azerbaijan\\
% ...
$^{85}$ CPPM, Aix-Marseille Universit{\'e} and CNRS/IN2P3, Marseille, France\\
% ...
$^{90}$ Department of Physics and Astronomy, Michigan State University, East Lansing MI, United States of America\\
% ...
$^{107}$ Nikhef National Institute for Subatomic Physics and University of Amsterdam, Amsterdam, Netherlands\\
% ...
$^{113}$ Homer L. Dodge Department of Physics and Astronomy, University of Oklahoma, Norman OK, United States of America\\
% ...
$^{116}$ Center for High Energy Physics, University of Oregon, Eugene OR, United States of America\\
% ...
$^{151}$ Institute of Physics, Academia Sinica, Taipei, Taiwan\\
$^{152}$ Department of Physics, Technion: Israel Institute of Technology, Haifa, Israel\\
$^{153}$ Raymond and Beverly Sackler School of Physics and Astronomy, Tel Aviv University, Tel Aviv, Israel\\
% ...
$^{158}$ Department of Physics, University of Toronto, Toronto ON, Canada\\
% ...
\end{flushleft}
